\documentclass[12pt]{article}
\usepackage{graphicx,  amsmath,amssymb,latexsym,psfrag,epsfig,subfigure,euscript, multirow}
\allowdisplaybreaks

\def\nb{{\bar n}}

\def\rd{\mathrm{d}}
\def\lep{{\sc lep} }
\def\lepone{{\sc lep} 1 }
\def\leptwo{{\sc lep} 2 }
\def\first{{1$^{\mathrm{st}}$}}
\def\second{{2$^{\mathrm{nd}}$}}
\def\third{{3$^{\mathrm{rd}}$}}
\def\fourth{{4$^{\mathrm{th}}$}}
\begin{document}

\begin{titlepage}

\begin{flushright}
FERMILAB-PUB-08-048-T\\
\end{flushright}

\vspace{0.2cm}
\begin{center}
\Large\bf
A precise determination of $\alpha_s$ from LEP thrust data using effective field theory
\end{center}

\vspace{0.2cm}
\begin{center}
{\sc Thomas Becher$^a$ and Matthew D. Schwartz$^{b}$}\\
\vspace{0.4cm}
{\sl $^a$\,Fermi National Accelerator Laboratory\\
P.O. Box 500, Batavia, IL 60510, U.S.A.\\[0.3cm]
$^b$\,Department of Physics and Astronomy
Johns Hopkins University\\
Baltimore, MD 21218, U.S.A.}
\end{center}

\vspace{0.2cm}
\begin{abstract}
\vspace{0.2cm}
\noindent 
Starting from a factorization theorem in Soft-Collinear Effective Theory, 
the thrust distribution in $e^+e^-$ collisions
is calculated including resummation of the next-to-next-to-next-to leading logarithms. 
This is a significant improvement over previous calculations
which were only valid to next-to-leading logarithmic order. 
The fixed-order expansion of the resummed result approaches the
exact fixed-order distribution towards the kinematic endpoint. 
This close agreement provides a verification of both the effective field theory
expression and recently completed next-to-next-to-leading fixed-order event shapes.
 The resummed distribution is then matched to
fixed order, resulting in a distribution valid over a large range of thrust. 
A fit to {\sc aleph} and {\sc opal} data from \lepone and \leptwo produces
$\alpha_s(m_Z)= 0.1172 \pm 0.0010 \pm 0.0008 \pm 0.0012 \pm 0.0012$, 
where the uncertainties are respectively statistical, systematic, hadronic, and perturbative.
This is one of the world's most precise extractions of $\alpha_s$ to date.
\end{abstract}
\vfil

\end{titlepage}

\section{Introduction}

Lepton colliders, such as the Large Electron-Positron collider \lep 
 which ran from 1989-2000 at {\sc{cern}}, provide an optimal 
environment for precision studies in high energy physics. Lacking the
complications of strongly interacting initial states, which plague hadron colliders, \lep has been able to provide extremely accurate measurements of standard model quantities such as the $Z$-boson mass, and its results tightly constrain beyond-the-standard model physics. 
The precision \lep data is also used for QCD studies, for example to determine the strong coupling constant $\alpha_s$.
With the variation of $\alpha_s$ known to 4-loops, one should be able to confirm
in great detail the running of the coupling, or use it to establish a discrepancy which might indicate new physics.
Even at fixed center-of-mass energy, differential distributions for event shapes, 
such as thrust probe several energy scales and are extremely sensitive to the running coupling.
 Moreover, event shape variables are designed to be infrared safe, so that they can be calculated in perturbation theory and so the theoretical
predictions should be correspondingly clean. 
Nevertheless, extractions of
$\alpha_s$ from event shapes at \lep have until now been limited by
theoretical uncertainty from unknown higher order terms in the perturbative expansion.

One difficulty in achieving an accurate theoretical prediction from QCD has
been the complexity of the relevant fixed-order calculations. Indeed, while
the next-to-leading-order (NLO) results for event shapes have been known since
1980~\cite{Ellis:1980wv},
the relevant next-to-next-to-leading order (NNLO) calculations were
completed only in 2007 \cite{GehrmannDeRidder:2007bj,GehrmannDeRidder:2007hr}.  
In addition to the loop integrals, the subtraction of soft and collinear divergencies in the real emission diagrams presented a major complication. 
In fact,
 this is the first calculation where a subtraction scheme has been successfully implemented at NNLO \cite{GehrmannDeRidder:2005cm}. 
However, even with these new results at hand,
the corresponding extraction of $\alpha_s$ continues to be limited by perturbative
uncertainty. The result of~\cite{Dissertori:2007xa} 
was $\alpha_s (m_Z) = 0.1240 \pm 0.0033$, with a perturbative uncertainty of 0.0029. This NNLO result for the strong coupling constant comes out lower than at NLO, but $2 \sigma$ higher  than the PDG average  $\alpha_s (m_Z)
= 0.1176 \pm 0.0020$ \cite{Yao:2006px}.
Actually, 
the most precise values of $\alpha_s$ are currently determined not from \lep but at low energies
using lattice simulations \cite{Mason:2005zx} and $\tau$-decays \cite{Davier:2005xq}. An extensive review of $\alpha_s$ determinations is given in \cite{Bethke:2006ac}, new determinations since its publication include \cite{Blumlein:2006be, Brambilla:2007cz}.

To further reduce the theoretical uncertainty of event shape calculations, 
it is important to resum the dominant perturbative
contributions to all orders in $\alpha_s$.
To see this, consider thrust, which is defined as
\begin{equation}
  T = \max_{\mathbf{n}} \frac{\sum_i | \mathbf{p}_i \cdot \mathbf{n}
  |}{\sum_i | \mathbf{\mathbf{p}}_i |}\,,
\end{equation}
where the sum is over all momentum 3-vectors $\mathbf{p}_i$ in the event,
and the maximum is over all unit 3-vectors ${\mathbf{n}}$. In the endpoint region, $T
\rightarrow 1$ or $\tau = (1 - T) \rightarrow 0$, no fixed-order
calculation could possibly describe the full distribution due to the
appearance of large logarithms. For example, at leading order in perturbation theory the thrust distribution has the form
\begin{equation}
 \frac{1}{\sigma_0}
\frac{\rd\sigma}{\rd\tau}
=\delta(\tau)+\frac{2\alpha_s}{3\pi} \left[\frac{-4\ln\tau-3}{\tau} + \dots\right] \,,
\end{equation}
where the ellipsis denotes terms that are regular in the limit $\tau\to 0$. Upon integration over the endpoint region, one finds
\begin{equation}
R(\tau)=\int_0^\tau \! \rd\tau' 
\frac{1}{\sigma_0}
\frac{\rd\sigma}{\rd\tau'}  
=1+ \frac{2\alpha_s}{3\pi} \left[-2\ln^2\tau-3\ln\tau + \dots\right] \,.
\end{equation}
Double logarithmic terms of the form $\alpha^n_s \ln^{2n}\tau$ arise from regions of
phase space where the quarks or gluons are soft or collinear.
For small enough $\tau$, higher order terms are just as
important as lower order ones and the standard perturbative expansion breaks down. Resummation refers to summing a series of contributions of the form $\alpha_s^n \ln^{m} \tau$ for the integral $R(\tau)$ or $\alpha_s^n (\ln^{m-1} \tau)/\tau$ for the differential distribution. Leading logarithmic (LL) accuracy is achieved by summing the tower of logarithms with $m=2n$, next-to-leading logarithmic accuracy (NLL) also sums the terms with $m=2n-1$. Resummation at N$^k$LL accuracy, provides all logarithmic terms with $2n \geq m \geq 2n-2k+1$, as detailed in Section \ref{sec:scet}.

The first resummation of event shapes was done by Catani, Trentadue, Turnock
and Webber (CTTW) in~\cite{Catani:1992ua}. 
Their approach was to define jet functions $J_C (p^2)$ as the probability for finding a jet of invariant mass $p^2$ in the event.
 These can be calculated to NLL by summing probabilities for successive emissions using the Alterelli-Parisi splitting functions.
 Each term in the series that is resummed corresponds to an additional semi-classical radiation.
 The splitting functions only account for collinear emissions;
to include soft emission, it is common either to impose some kind of angular ordering constraint to
simulate soft coherence effects, or to use more sophisticated probability functions, such as Catani-Seymour dipoles~\cite{Catani:1996vz}.
Except for \cite{deFlorian:2004mp}, none of these approaches has led to a resummation for event shapes beyond NLL.

The approach to resummation of event shapes~\cite{Schwartz:2007ib} based 
on Soft-Collinear Effective Theory 
(SCET)~\cite{Bauer:2000yr,Bauer:2001yt,Beneke:2002ph} 
contrasts sharply with the semi-classical CTTW treatment. The most important conceptual difference is
that effective field theory works with amplitudes, at the operator level,
instead of probabilities at the level of a differential cross-section.
Consequently, the resummation comes not from the exponentially decreasing
probability for multiple emissions, but from a solution to renormalization
group (RG) equations. 

The starting point for the effective field theory approach is the factorization
formula for thrust in the 2-jet region,
\begin{equation}
  \frac{1}{\sigma_0} \frac{\rd \sigma_{2}}{\rd \tau} = H(Q^2, \mu) \int \! 
\rd  p^2_L \rd p_R^2  \rd k \,J (p^2_L, \mu) \,J (p^2_R, \mu)\, S_T (k, \mu) \delta (\tau -
  \frac{p^2_L+p^2_R}{Q^2}-\frac{k}{Q}) \label{scetfact}\,,
\end{equation}
where $H (Q^2, \mu)$ is the hard function, $J(p^2, \mu)$ the jet function,
and $S_T (k, \mu)$ is the soft function for thrust. $Q$ refers to the
center-of-mass energy of the collision, $\mu$ is an arbitrary
renormalization scale, and the born-level cross section $\sigma_0$ appears for normalization. 
A similar factorization formula was derived to study top quark jets
in~\cite{Fleming:2007qr}, and then transformed into this form to study 
event shapes in~\cite{Schwartz:2007ib}. 
Factorization properties of event shape variables were also studied in~\cite{Berger:2003gr,Berger:2003pk}.
The expression (\ref{scetfact}) is valid to all orders in perturbation 
theory up to terms which are power suppressed in the two-jet region $\tau\rightarrow 0$, 
\begin{equation}
 \frac{\rd \sigma}{\rd \tau} =  \frac{\rd \sigma_2}{\rd \tau} \Big[1+{\mathcal O}(\tau)\Big]\,.
\end{equation}

The key to the factorization theorem is that near maximum thrust, $\tau$ reduces to
the sum of hemisphere masses
\begin{equation}
  \tau \to \frac{M_L^2+M_R^2}{Q^2}=\frac{p_L^2+p_R^2+k Q}{Q^2}\,,
\end{equation}
where the two hemispheres are
defined by the thrust axis $\mathbf{n}$. Here, $p_L^2 (p_R^2)$ is the invariant mass of the energetic particles in the left (right) jet and $k Q$ is the increase of the invariant mass on the two sides due to soft emissions.  A more detailed interpretation of this formula can be found in~\cite{Schwartz:2007ib,Fleming:2007qr,Fleming:2007xt}. 

The factorization theorem (\ref{scetfact}) makes it evident that the thrust distribution involves three different scales in the endpoint region. First of all, there are virtual effects arising in the production of the quark anti-quark pair at the hard scale $\mu_h\sim Q$ which are encoded in the hard function $H(Q^2,\mu)$.
 A second relevant scale is associated with the invariant mass of the two back-to-back jets,
 $\mu_j^2\sim p_L^2+p_R^2 \sim \tau Q^2$. 
In addition to these two external scales, a third, lower seesaw scale is 
encoded in the soft function 
$\mu_s \sim k\sim \tau Q \sim \mu_j^2/\mu_h$. 
The effective theory treatment separates the effects associated with these three scales and makes
transparent that a larger range of scales, and consequently a larger range of
$\alpha_s (\mu)$ is being probed than is evident in either the fixed-order
calculation or in the traditional NLL resummation.
Large logarithms are avoided using RG evolution in the effective theory.
Each of the three functions $H$, $J$ and $S$ is evaluated at its characteristic scale,
and then evolved to a common scale $\mu$.
Solving the differential RG equations resums logarithms of the scale ratios. 

In Section \ref{sec:scet}, we provide the definitions of $H$, $J$ and $S$ in SCET. 
These functions can be calculated directly in SCET,
or one can rewrite them in terms of matrix elements of QCD operators. 
For practical calculations, the definitions in QCD are often more suitable since the QCD Feynman rules are simpler. The hard and jet function appear in other processes and are known to two-loop order \cite{Becher:2006qw, Becher:2006mr,Becher:2007ty}. 
Their RG-equations have been solved in closed form and the relevant anomalous dimensions are known at three-loop order \cite{Becher:2006nr}.
With the hard and jet functions known, the only missing ingredient to resum the thrust distribution to next-to-next-to-next-to-leading logarithmic (N$^3$LL) accuracy is the soft function $S$. Its one-loop expression was given in \cite{Schwartz:2007ib} and in Section \ref{sec:match} we determine soft function to two loops. Its logarithmic part is obtained using RG invariance of the thrust distribution and the remaining constant piece by a numerical procedure.
After plugging the solutions back into the factorization theorem (\ref{scetfact}), we obtain the result for the resummed distribution valid to N$^3$LL.

Next, we expand the effective theory result to fixed order in $\alpha_s$.
The logarithmically enhanced terms which are determined in the effective theory
dominate the thrust distribution. This is especially pronounced at NNLO: color structure by color structure, we find that the logarithmic terms are an excellent approximation of the full fixed-order result. This close agreement also provides an independent check on the NNLO calculation.
After comparing the full fixed-order result to the logarithmic terms, we add the small difference between the two to our resummed result. By this matching procedure, we obtain a resummed result which is also correct to NNLO in fixed-order perturbation theory.

In Section \ref{sec:fit}, we fit the resummed matched calculation to {\sc{aleph}} and {\sc{opal}} data. We find 
a relatively small perturbative uncertainty on $\alpha_s$ compared to previous event shape fits of the same \lep
data. In fact, the final statistical, systematic, perturbative and hadronization uncertainties end up being quite similar,
all around 1\%. At this point, we have the least handle on hadronization effects, and these and other
power corrections are explored in Section \ref{sec:NP}. The conclusion contains a brief
discussion of how the various uncertainties might be further reduced.

\section{Resummation of thrust in effective field theory \label{sec:scet}}

The large logarithms in the thrust distribution dominate near the endpoint,
$\tau \rightarrow 0$. This region of phase space corresponds to configurations
with two back-to-back light jets. In this situation, the vector and axial-vector currents relevant to the production of the $q\bar q$-pair are
mapped onto the two-jet operators in SCET~\cite{Bauer:2002nz}
\begin{equation}
  \mathcal{O}_2 = \bar\chi_{\bar{n}} \Gamma \chi_n\,,
\end{equation}
where $\Gamma=\gamma^\mu$ or $\Gamma=\gamma^\mu \gamma_5$ for vector or axial-vector currents respectively.
Here, $n$ is a light-like 4-vector aligned with the thrust
axis and the composite 
fields $\chi_n$ and $\chi_{\bar{n}}$ are the collinear quark fields in the $n$- and $\bar n$-directions, multiplied by light-like Wilson lines~\cite{Hill:2002vw}.
The first step in the effective field theory calculation is matching to full
QCD. This is done by calculating matrix elements in SCET and in QCD and adjusting the Wilson coefficients 
in the effective theory so that the matrix elements agree. Performing the matching on-shell, one finds that the relevant matching coefficient for the vector operator is given by the on-shell
vector quark form factor. In a scheme with an anti-commuting $\gamma_5$, the Wilson coefficients of the
vector and axial-vector operators are identical. Neglecting electro-weak corrections, the use of such a scheme is consistent in the endpoint region $\tau\to0$. 
In this region, the two energetic quarks produced directly by the current always appear in the final state, 
so that the $\gamma_5$ matrices from the axial currents always appear in a single trace formed by the cut fermion loop.

After normalizing to the tree-level cross section, the hard function $H(Q,\mu)$ is given by the absolute value squared of the time-like on-shell form factor. Using the known two-loop result for the on-shell QCD form factor \cite{Matsuura:1987wt,Matsuura:1988sm,Gehrmann:2005pd,Moch:2005id}, the hard function at two loops was derived in \cite{Becher:2006mr}. It satisfies the RG equation \cite{Becher:2006nr}
\begin{eqnarray}\label{Hrge}
   \frac{\rd}{\rd\ln\mu}\,H(Q^2,\mu)
   &=& \left[ 2\Gamma_{\rm cusp}(\alpha_s)
   \ln\frac{Q^2}{\mu^2} 
   + 2\gamma^H(\alpha_s) \right] H(Q^2,\mu) \,, 
\end{eqnarray}
whose solution can be written as
\begin{equation}
  H(Q^2, \mu) =  H(Q^2, \mu_h)  \exp \left[ 4 S (\mu_h,
  \mu) - 2A_H (\mu_h, \mu) \right]\, \left( \frac{Q^2}{\mu_h^2} \right)^{-2A_\Gamma(\mu_h,\mu)}\,.
\end{equation}
Here, 
\begin{equation}\label{Sfun}
  S (\nu, \mu) = - \int_{\alpha_s (\nu)}^{\alpha_s (\mu)} \rd \alpha
  \frac{\Gamma_{\mathrm{cusp}} (\alpha)}{\beta (\alpha)} \int_{\alpha_s
  (\nu)}^{\alpha} \frac{\rd \alpha'}{\beta (\alpha')}
\end{equation}
and
\begin{equation}\label{Afun}
  A_H (\nu, \mu) = - \int_{\alpha_s (\nu)}^{\alpha_s (\mu)} d \alpha
  \frac{\gamma^H (\alpha)}{\beta (\alpha)}\,.
\end{equation}
The function $A_\Gamma(\nu,\mu)$ is defined as $A_H (\nu, \mu)$, but with $\gamma^H$ replaced by  $\Gamma_{\rm cusp}$. The solutions of the RG equations for the jet and soft function given below involve functions $A_J (\nu,\mu)$, $A_S (\nu, \mu)$ which are obtained from (\ref{Afun}) by substituting $\gamma_J, \gamma_S$ for $\gamma_H$.
It is straightforward to expand $S (\nu, \mu)$ and $A_H (\nu, \mu)$
perturbatively in $\alpha_s (\nu)$ and $\alpha_s (\mu)$ given the expansions
of $\Gamma_{\mathrm{cusp}} (\alpha)$ and $\gamma^H (\alpha)$. The explicit 
expansions can be found in~\cite{Becher:2006mr}.

In SCET the jet function is given by the imaginary part of the collinear quark propagator,
\begin{equation}
  J (p^2, \mu) =  \frac{1}{( \bar{n} \cdot p)} \frac{1}{\pi} \mathrm{Im} \left[ i \int d^4x\,
   e^{- i p x} \langle 0 |  {\bf T} \left\{ \bar\chi_n(x)\,\frac{\nb\!\!\!/}{2}\,  {\chi}_{n} (0)
   \right\} |0 \rangle \right]
=\delta(p^2) + {\cal O}(\alpha_s) 
\,
\end{equation}
and thus vanishes for $p^2<0$. The jet function was calculated at one loop in~\cite{Bauer:2003pi} and at two loops in~\cite{Becher:2006qw}. To evaluate the function perturbatively, it is convenient to rewrite the collinear quark propagator in terms of QCD fields. One finds that the jet function is obtained from the quark propagator in light-cone gauge. The jet function satisfies a RG
equation which is non-local in $p^2$~\cite{Becher:2006qw},
\begin{equation}
  \frac{\rd J (p^2, \mu)}{\rd \ln \mu} = \left[ - 2 \Gamma_{\mathrm{cusp}} \ln
  \frac{p^2}{\mu^2} - 2 \gamma_J \right] J (p^2, \mu) + 2 \Gamma_{\mathrm{cusp}}
  \int_0^{p^2} d q^2 \frac{J (p^2, \mu) - J (q^2, \mu)}{p^2 - q^2}\,.
\end{equation}
From the divergent part of the form factor at three
loops~\cite{Moch:2005id} 
and the NNLO Altarelli-Parisi splitting functions~\cite{Moch:2004pa} 
the jet anomalous dimension $\gamma_J$ was derived at three loops
in~\cite{Becher:2006mr} and is given in Appendix \ref{sec:anomalous}. Although the RG equation is non-local in $p^2$, it is local in $\mu$ and can be solved using Laplace transform techniques. The result is~\cite{Becher:2006mr}
\begin{equation}\label{jetsol}
  J (p^2, \mu) = \exp \left[ - 4 S (\mu_j, \mu) + 2 A_J (\mu_j, \mu) \right]
  \widetilde{j} (\partial_{\eta_j},\mu_j)  \frac{1}{p^2} \left(
  \frac{p^2}{\mu_j^2} \right)^{\eta_j}  \frac{e^{- \gamma_E
  \eta_j}}{\Gamma (\eta_j)} \,,
\end{equation}
where $\eta_j = 2 A_{\Gamma} (\mu_j, \mu)$. The function $\widetilde j(L,\mu)$ is the Laplace transform of the jet function. Its definition and explicit form are given in Appendix \ref{sec:Hjs}. To any given order in perturbation theory, $\widetilde j(L,\mu)$ is a polynomial in the variable $L$ so that the derivatives with respect to $\eta_j$ in (\ref{jetsol}) can be performed explicitly. 

The thrust soft function is defined as a matrix element of Wilson
lines along the directions of the energetic quarks,
\begin{equation}
  S_T (k) = \sum_X \left| \left\langle X|Y_n^{\dag} Y_{\bar{n}} |0  \right\rangle \right|^2
\delta (k - n \cdot p_{X_n} - \bar{n} \cdot p_{X_{\bar{n}}})\,,
\end{equation}
where 
\begin{equation}\label{eq:Sn}
   Y_n = {\rm\bf P}\,\exp\left(
   ig\int_{-\infty}^0\!\rd t\,n\cdot A_{s}(t n) \right)\,.
\end{equation}
This Wilson line describes the Eikonal interactions of soft gluons with the fast moving quark, and $p_{X_n}$ ($p_{X_\nb}$) is the sum of the momenta of the soft particles in the $n$-hemisphere ($\nb$-hemisphere). The variable $k$ measures the change in the invariant mass due to soft emissions from the two jets.  At the leading power, the mass in the $n$-hemisphere is given by
\begin{eqnarray}
M_n^2  &=& (p_n+p_{X_n})^2 \approx p_n^2+ Q\,( n \cdot p_{X_n})   \,, 
\end{eqnarray}
where $p_n$  denotes the total collinear momentum in the hemisphere. Note that the soft function vanishes for negative argument.

Like the jet function, the soft function can be calculated
order-by-order in perturbation theory. The one-loop
soft function was derived in~\cite{Schwartz:2007ib}
 from results of~\cite{Korchemsky:1993uz}; it was also calculated
directly in SCET~\cite{Fleming:2007xt}. 
The two-loop soft function will be determined below. 

The factorization theorem (\ref{scetfact}) and the fact that the thrust distribution is independent of the renormalization scale $\mu$ implies that the soft function fulfills the RG equation
\begin{equation}\label{Srge}
   \frac{{\rm d}S_T(k,\mu)}{{\rm d}\ln\mu}
   = \left[ 4\Gamma_{\rm cusp}(\alpha_s)\,\ln\frac{k}{\mu}
    - 2\gamma^S(\alpha_s) \right] S_T(k,\mu) 
   \mbox{}-4\Gamma_{\rm cusp}(\alpha_s) \int_0^{k}\!dk'\,
    \frac{S_T(k,\mu)-S_T(k',\mu)}%
         {k-k'} \,,
\end{equation} 
and that, to all orders,
\begin{equation}\label{gw}
  \gamma^S = \gamma^H- 2 \gamma^J   \,.
\end{equation}
This relation was checked to one loop in~\cite{Schwartz:2007ib} (with a different convention for $\gamma_H$), and here we use it to determine the two- and three-loop soft anomalous dimensions. 
Similar to (\ref{jetsol}), the solution for the soft function is
\begin{equation}\label{softsol}
  S_T(k, \mu) = \exp \left[ 4 S (\mu_s, \mu) + 2 A_S (\mu_s, \mu) \right]
  \widetilde{s}_T (\partial_{\eta_s})  \frac{1}{k} \left( \frac{k}{\mu_s}
  \right)^{\eta_s}  \frac{e^{- \gamma_E \eta_s}}{\Gamma \left( \eta_s \right)} \,,
\end{equation}
with $\eta_s = - 4 A_{\Gamma} (\mu_s, \mu)$.
From the linearity of $A_S$ in
$\gamma_S$ it also follows that $A_S = A_H - 2 A_J$.

The convolution integrals in (\ref{scetfact}) can be done analytically once the solutions
  (\ref{jetsol}) and (\ref{softsol}) are put back into the factorization theorem.
 The thrust distribution becomes
\begin{multline}\label{complicated}
  \frac{1}{\sigma_0} \frac{{\rm d} \sigma_2}{{\rm d} \tau} = \exp \left[ 4 S (\mu_h, \mu)
  - 2 A_H (\mu_h, \mu) - 8 S (\mu_j, \mu) + 4 A_J (\mu_j, \mu) + 4 S (\mu_s,
  \mu) + 2 A_S (\mu_s, \mu) \right] \\
  \times \left(\frac{Q^2}{\mu_h^2}\right)^{-2A_\Gamma(\mu_h,\mu)}  H(Q^2,\mu_h) \left[ \widetilde{j}
  (\partial_{2 \eta_j},\mu_j)\right]^2 \widetilde{s}_T (\partial_{\eta_s},\mu_s) \left[
  \frac{1}{\tau} \left( \frac{\tau Q^2}{\mu_j^2} \right)^{2 \eta_j} \left(
  \frac{\tau Q}{\mu_s} \right)^{\eta_s} \frac{e^{- \gamma_E (2 \eta_j +
  \eta_s)}}{\Gamma (2 \eta_j + \eta_s)} \right] \,.
\end{multline}
Using the relations,
\begin{align}
 A_\Gamma(\mu_1,\mu_2) +A_\Gamma(\mu_2,\mu_3) &= A_\Gamma(\mu_1,\mu_3)\,, \\
 S (\mu_1, \mu_2) + S(\mu_2, \mu_3) &= S (\mu_1, \mu_3) + \ln \frac{\mu_1}{\mu_2} A_{\Gamma} (\mu_2, \mu_3) \,, \nonumber
\end{align}
and
\begin{equation}
  f (\partial_{\eta}) X^{\eta}  = X^{\eta} f (\ln X +
  \partial_{\eta})\,,
  \end{equation}
the expression (\ref{complicated}) simplifies to
\begin{multline} \label{scetdist}
\frac{1}{\sigma_0} \frac{\rd\sigma_2}{\rd\tau}=
\exp\left[ 4S(\mu_h,\mu_j)+4S(\mu_s,\mu_j)-2A_H(\mu_h,\mu_s)+4A_J(\mu_j,\mu_s)\right] \left(\frac{Q^2}{\mu_h^2}\right)^{-2A_\Gamma(\mu_h,\mu_j)} \\
\times H(Q^2,\mu_h)\, 
\left[\widetilde j\Big( \ln\frac{\mu_s Q}{\mu_j^2}+\partial_\eta,\mu_j\Big)\right]^2\, \widetilde s_T\Big(\partial_\eta,\mu_s\Big) \frac{1}{\tau} \left(\frac{\tau Q}{\mu_s}  \right)^{\eta} \frac{e^{-\gamma_E \eta}}{\Gamma(\eta)}\,,
\end{multline}
with $\eta = 4 A_{\Gamma} (\mu_j, \mu_s)$. From this final result we can read off the canonical relations among the
hard, jet, and soft matching scales and the physical scales $Q$ and $p \sim
\sqrt{\tau} Q$:
\begin{equation}\label{canonical}
  \mu_h = Q\,, \hspace{1em} \mu_j = \sqrt{\tau} Q\,, \hspace{1em} \mu_s = \tau Q\,.
\end{equation}
Note that the arbitrary reference scale $\mu$ has dropped out completely.

For the $\alpha_s$ fits, we need  the differential thrust distribution integrated over each bin.
The integral of the thrust distribution can be evaluated analytically, since the derivatives with respect to $\eta$  in (\ref{scetdist}) commute with the integration over $\tau$. The resulting expression is  
\begin{multline}
  R_2(\tau) = \int_0^{\tau} \frac{1}{\sigma_0} \frac{\rd \sigma_2}{\rd \tau'} \rd \tau'= \exp \left[4 S (\mu_h, \mu_j) + 4 S (\mu_s, \mu_j) - 2 A_H
  (\mu_h, \mu_s) + 4 A_J (\mu_j, \mu_s) \right] \\
  \times  \left(\frac{Q^2}{\mu_h^2}\right)^{-2A_\Gamma(\mu_h,\mu_j)} 
H(Q^2,\mu_h) \left[ \widetilde{j} (\ln
  \frac{\mu_s Q}{\mu_j^2} + \partial_{\eta},\mu_j) \right]^2 \widetilde{s}_T
  (\partial_{\eta},\mu_s) \left[ \left( \frac{\tau Q}{\mu_s} \right)^{\eta}
  \frac{e^{- \gamma_E \eta}}{\Gamma (\eta + 1)} \right] \,. \label{R2eq}
\end{multline}
Note that the integral is performed for fixed $\mu_j$ and $\mu_s$,
that is, before setting them to their canonical $\tau$-dependent values. In this way, large logarithms are removed in the observable of interest, not for some intermediate expression.

\begin{table}
\begin{center}
  \begin{tabular}{|c|c|c|c|c|c|c|}  \hline
   \multirow{2}{*}{order}   &     \multirow{2}{*}{$\Gamma_{\rm cusp}$} &     \multirow{2}{*}{$\gamma^{H/J/S}$} &     \multirow{2}{*}{$H$, $\widetilde j$, $\widetilde s_T$} &     \multirow{2}{*}{$\beta$} & fixed-order  & logarithmic \\   
     &   &            &        &    &  matching  &  accuracy    \\  \hline
    \first order    &  2-loop  &  1-loop          &    tree     &   2-loop&   -- & NLL      \\
    \hline
    \second order  &  3-loop  &  2-loop          &    1-loop   &   3-loop&   LO   & NNLL  \\
    \hline
    \third order    &  4-loop  &  3-loop          &    2-loop   &   4-loop&   NLO  & N$^3$LL  \\
    \hline
    \fourth order   &  4-loop  &  3-loop          &    3-loop   &   4-loop&   NNLO & N$^3$LL  \\
    \hline
\end{tabular}
\end{center}
\caption{Definition of orders in perturbation theory}
\label{tab:ords}
\end{table}

Different definitions of logarithmic accuracy are commonly used in the literature. Before proceeding further, we now show which logarithms are included at a given order in our calculation. We use renormalization-group improved perturbation theory, in which logarithms of scales are eliminated in favor of coupling constants at different scales which are counted as small parameters of the same order
\begin{equation}
\ln \frac{\mu}{\nu} = \int_{\alpha_s(\nu)}^{\alpha_s(\mu)} \frac{d\alpha}{\beta(\alpha)} = \frac{2\pi}{\beta_0}\left(\frac{1}{\alpha_s(\mu)}-\frac{1}{\alpha_s(\nu)}\right)
+\dots\,.
\end{equation}
The expansion of the Sudakov exponent (\ref{Sfun}) then takes the form
\begin{equation} \label{RGI}
S(\nu,\mu)= \frac{1}{\alpha_s(\nu)} f_1(r) + f_2(r) + \alpha_s(\nu) f_3(r)  + \alpha_s(\nu)^2 f_4(r)+ \dots
\end{equation}
where $r=\alpha_s(\mu)/\alpha_s(\nu)$. The explicit expressions for the functions $f_1$ to $f_4$ needed for our calculation are given in \cite{Becher:2006mr}. The leading-order term $\alpha_s^0$ in renormalization group improved perturbation theory involves the functions $f_1$ and $f_2$, which depend on the one and two-loop cusp anomalous dimension. To make contact with the literature, we can expand $\alpha_s(\mu)$ around fixed coupling $\alpha_s\equiv\alpha_s(\nu)$. The result takes the form
\begin{equation} \label{LOG}
S(\nu,\mu)= L\, g_1(\alpha_s L) + g_2(\alpha_s L) + \alpha_s g_3(\alpha_s L) +\alpha_s^2 g_4(\alpha_s L) + \dots\,,
\end{equation}
with $L=\ln(\mu/\nu)$. LL resummations include only $g_1$, NLL also $g_2$ and so forth. When rewriting (\ref{RGI}) in the form (\ref{LOG}), the expansion of $f_i$ contributes to the functions $g_j$ with $j\geq i$ so that there is a one-to-one correspondence between the order in renormalization group improved perturbation theory and the standard logarithmic accuracy. 
Note that the higher order terms to (\ref{RGI}) and (\ref{LOG}) are suppressed by explicit factors of $\alpha_s$. The missing pieces in the integral $R_2(\tau)$ at N$^3$LL are suppressed by $\alpha_s^3$ so that the missing logarithms are $\alpha^3\times\alpha^{n} \ln^{2n}\tau \equiv \alpha^k \ln^{2k-6}\tau$  for the default scale choice. In particular, at order $\alpha^3$ the N$^3$LL result includes everything except for the constant term in $R_2(\tau)$ which does not contribute to the thrust distribution. 

In Table \ref{tab:ords}, we list the ingredients to obtain $(\ref{R2eq})$ to a given accuracy. The necessary anomalous dimensions and the results for the functions $H$, $\widetilde{j}$ and $\widetilde{s}_T$ are provided in Appendix \ref{sec:anomalous}. Everything in the table except for the four-loop cusp anomalous dimension and the constant part of the two-loop soft function are known. We estimate the former using the Pad\'e approximation $\Gamma_4=\Gamma_3^2/\Gamma_2$~\cite{Moch:2005ba} and determine the latter numerically in the next section. Rather than specifying both the accuracy of the resummation and the order to which we match to the fixed order result, will will in the following simply refer to the definitions of \first, \second, \third  and \fourth order 
as given in Table \ref{tab:ords}. Note that the difference between \third and \fourth order, as we have defined them, is only
the inclusion of NNLO matching corrections, but the logarithmic accuracy stays the same.

\section{Resummation vs. fixed order \label{sec:match}}

\begin{figure}[t!]
\begin{center}
\includegraphics[width=0.95\textwidth]{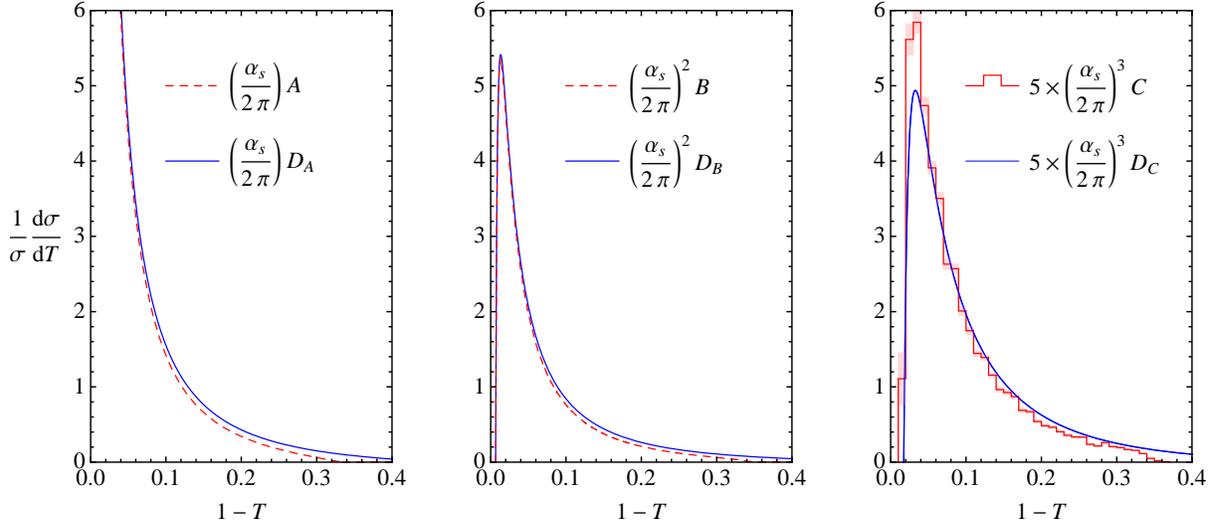}
\end{center}
\vspace*{-1.2cm}
\caption{A comparison of the full fixed-order calculations and
the fixed-order expansion of the resummed distributions from the effective field theory. The light-red areas in the NNLO histogram are an estimate of the statistical uncertainty.}
\label{fig:ABC}
\end{figure}

In this section, we compare the resummed expression, valid in the endpoint
region $\tau \to 0$ to the fixed-order expression, which is valid away from
the endpoint. The resummed expression, when expanded to fixed order, must
reproduce the $\tau = 0$ singularities of the fixed-order calculation. This
observation can be used to extract numerically the constant part of the two-loop
soft function. Then, by including the difference between the expanded resummed expression and the fixed-order expression, we derive the final matched distribution.

The fixed-order thrust distribution has been calculated to leading order
analytically and to NLO and NNLO
numerically. For the scale choice $\mu=Q$, the result is usually written in the form
\begin{equation}\label{fixeddist}
 \frac{1}{\sigma_0} \frac{\rd \sigma}{\rd \tau}  =
  \delta (\tau) + \left( \frac{\alpha_s}{2 \pi} \right) A (\tau) + \left(
  \frac{\alpha_s}{2 \pi} \right)^2 B (\tau) + \left( \frac{\alpha_s}{2 \pi}
  \right)^2 C (\tau) + \cdots\,,
\end{equation}
where we have suppressed the argument of the coupling constant, $\alpha_s \equiv \alpha_s (Q)$. Throughout the following analysis, we use an analytical form for $A (\tau)$, a numerical calculation of $B (\tau)$ using 
the program {\sc{event2}}~\cite{Catani:1996jh} with $10^{10}$ events and a 
numerical calculation of $C (\tau)$ that was generously provided by the authors of~\cite{GehrmannDeRidder:2007bj}. A value of $y_0=10^{-5}$ for the infrared cut-off was used in the calculation of the NNLO histograms, see \cite{GehrmannDeRidder:2007jk}.

The resummed differential thrust distribution in the effective theory is given
in Eq. (\ref{scetdist}). To compare with fixed-order results (\ref{fixeddist}), 
we set all scales equal $\mu_h=\mu_j=\mu_s=Q$. Doing so switches off the resummation: 
all evolution factors, such as $S (\mu_h, \mu_j)$ and $A_H(\mu_h,\mu_s)$, vanish in the limit of equal scales. Before taking the limit $\eta =4A_\Gamma(\mu_j,\mu_s)\rightarrow 0$, we expand the kernel in (\ref{scetdist}) using the relation
\begin{equation}
\frac{1}{\tau^{1-\eta}}= \frac{1}{\eta} \delta(\tau) +\sum_{n=0}^{\infty} \frac{\eta^n}{n!} \left[\frac{\ln^n\tau}{\tau}\right]_+\,,
\end{equation}
and evaluate the derivatives with respect to $\eta$ using the explicit expressions for $\widetilde{j}$ and $\widetilde{s}$. The result is a
sum of distributions
\begin{equation}
  \frac{1}{\sigma_0} \frac{\rd \sigma_2}{\rd \tau} = \delta (\tau) D_{\delta} +
  \left( \frac{\alpha_s}{2 \pi} \right) \left[ D_A (\tau) \right]_+ + \left(
  \frac{\alpha_s}{2 \pi} \right)^2 \left[ D_B (\tau) \right]_+ + \left(
  \frac{\alpha_s}{2 \pi} \right)^3 \left[ D_C (\tau) \right]_+ + \cdots\,.
  \label{foscet}
\end{equation}
The coefficients $D_{\delta}$, $D_A$ $D_B$ and $D_C$ are given in Appendix~\ref{sec:singular}. Away from $\tau = 0$, the $\delta$-function terms can be dropped and the plus-distributions reduce to their argument functions, $[D_X (\tau)]_+ =D_X (\tau)$. Since the effective field theory resums the large logarithms of
the fixed-order distribution, there should not be any $1/\tau$
singularities in $A$, $B$, or $C$ which are not reproduced in $D_A$, $D_B$ and
$D_C$ respectively. This was shown analytically for the $A$ function 
in~\cite{Schwartz:2007ib}. It is demonstrated numerically for $A$, $B$ and $C$ in Figure \ref{fig:ABC}. In fact, the figure shows that even at moderate $\tau$, the thrust distribution is dominated by the singular terms. Note that the lowest three bins of the numerical result for $C$ are above the effective theory prediction. This is due to numerical difficulties in the fixed-order code used to evaluate $C$ and will be explored in more detail below.

\begin{figure}[t!]
\psfrag{x}[]{\small $1-T$}
\psfrag{y}[]{{\small $(1-T)$}{\large $\frac{1}{\sigma}$}{\Large$\frac{\rd \sigma}{\rd T}$}}
\begin{center}
\includegraphics[width=0.6\textwidth]{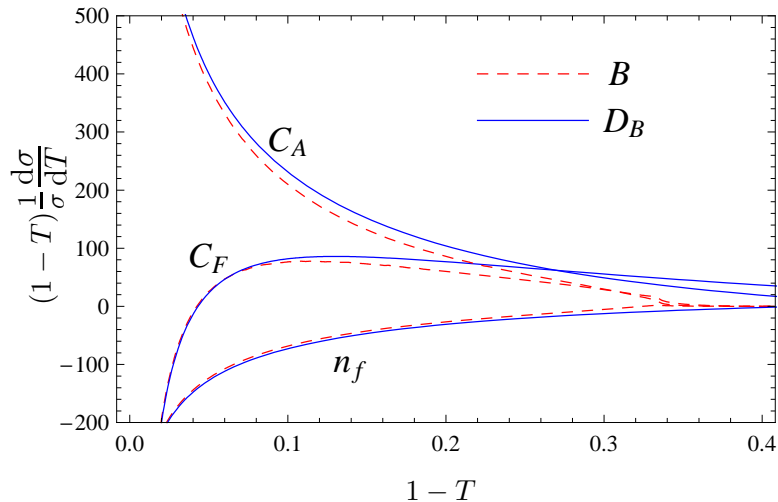}
\end{center}
\vspace*{-0.7cm}
\caption{Color structures used in NLO comparison with fixed order.}
\label{fig:NLOcols}
\end{figure}%

The SCET expression (\ref{scetdist}) for the thrust distribution is valid as
$\tau \rightarrow 0$, that is, in the 2-jet region. One could perform resummation also for terms which are power suppressed in this limit, by including operators with additional fields or derivatives into the effective theory~\cite{Bauer:2006mk,Bauer:2006qp}. However, since these terms are power suppressed it is sufficient to include them at fixed order. 
To do so, we simply subtract the singular terms from the fixed-order expression. The remainder is
\begin{multline}
 r(\tau)\equiv \frac{1}{\sigma_0}\left( \frac{\rd \sigma}{\rd \tau} -  \frac{\rd \sigma_2}{\rd \tau}\right) \\
 = \left( \frac{\alpha_s}{2
  \pi} \right) \left[ A (\tau) - D_A (\tau) \right]
  + \left( \frac{\alpha_s}{2 \pi} \right)^2 \left[ B (\tau) - D_B (\tau)
  \right] + \left( \frac{\alpha_s}{2 \pi} \right)^3 \left[ C (\tau) - D_C
  (\tau) \right] + \cdots \,.\label{scet3}
\end{multline}
Including the matching contribution, the thrust distribution becomes
\begin{equation}
  \frac{1}{\sigma_0}\frac{\rd \sigma}{\rd \tau} = \frac{1}{\sigma_0}\frac{\rd \sigma_2}{\rd \tau} + r(\tau) \,. \label{scet23}
\end{equation}
With the inclusion of $r(\tau)$, our result not only resums the thrust distribution to N$^3$LL, but is is also correct to NNLO in fixed-order perturbation theory.

\begin{figure}[t!]
\begin{center}
\psfrag{y}[]{$c_2^S$}
\includegraphics[width=0.6\textwidth]{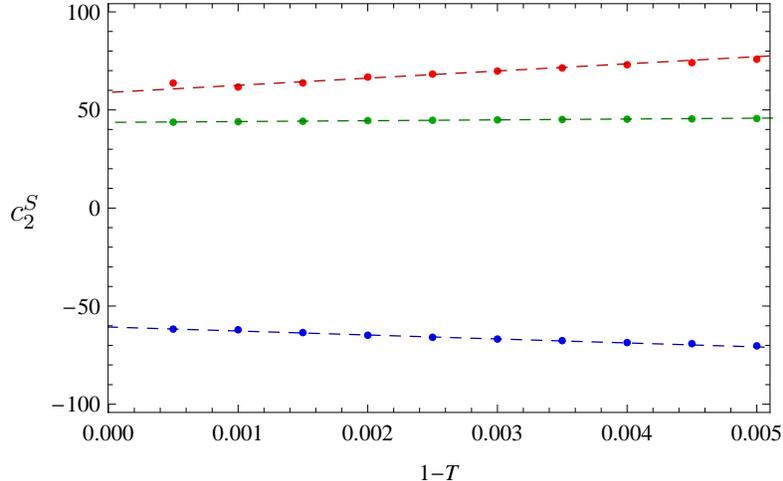}
\end{center}
\vspace*{-0.8cm}
\caption{Extraction of the two-loop constants in the soft function. The points correspond to the value of an infrared cutoff applied to the fixed-order calculation. The lines are
interpolations among the points from $\tau=0.001$ to $\tau=0.003$ 
extrapolated to $\tau=0$ to extract
the constants. From top to bottom, the curves are the $C_F^2, C_A$ and $n_f$ color factors.}
\label{fig:softNLO}
\end{figure}

Now let us turn to the two-loop soft function.
Its RG equation together with the anomalous dimensions determine
 the logarithmic part of ${\widetilde s}_T$, but the constant part 
\begin{equation}
{\widetilde s}_T(0,\mu) = 1 + C_F\frac{\alpha_s}{4\pi} \left( -\pi ^2 \right)
   + C_F \left( \frac{\alpha_s}{4\pi} \right)^2
   \left[  C_F\,c_{2,C_F}^S +  C_A\, c_{2,C_A}^S + T_F \,n_f \, c_{2,n_f}^S  \right] 
\end{equation}
cannot be obtained in this way. 
We will determine the constant from the requirement that the integral over the thrust distribution reproduces the total hadronic cross section
\begin{equation}
  \frac{\sigma_{\rm had}}{\sigma_0} = 1 + \frac{\alpha_s}{4 \pi} \left[ 3 C_F \right]
  + \left( \frac{\alpha_s}{4 \pi} \right)^2 \left[  C_F C_A \left(
  \frac{123}{2} - 44 \zeta_3 \right) +  C_F T_F n_f \left( - 22 + 16
  \zeta_3 \right) - C_F^2 \frac{3}{2} \right]\, . \label{sigtot}
\end{equation}
Plugging (\ref{foscet}) and
(\ref{scet3}) into ($\ref{scet23}$) we find
\begin{equation}
  \frac{\sigma_{\rm had}}{\sigma_0} = D_{\delta} + \int_0^1  d \tau\, r(\tau)
  = 1 + \frac{\alpha_s}{\pi}
  + \left( \frac{\alpha_s}{4 \pi} \right)^2 \left\{ 317.5 +
  c_{2}^S  + 4\int_0^1 \left[ B (\tau) - D_B (\tau) \right] d \tau
  \right\}\,,
\end{equation}
where 317.5 comes from setting $n_f=5$ in $D_\delta$ (see Appendix C).
Since we know separately the color structures for $B$ (numerically) and $D_B$ (analytically), as shown in Figure \ref{fig:NLOcols}, we can perform this integral numerically and then extract $c_2^S$ by comparing to (\ref{sigtot}). Although
the difference $B (\tau) - D_B (\tau)$ is integrable as $\tau \rightarrow 0$ both
of these functions are separately divergent. To have numerically stable results, we impose an infrared cutoff
$\tau_0$ on the integral and interpolate to $\tau_0 = 0$. We do this in
discrete steps by dropping the lowest bins in the $B (\tau)$ distribution
which was generated with the {\sc{event2}} program. The convergence and
interpolation are shown in Figure \ref{fig:softNLO}. We find
\begin{align}\label{softConst}
  c_{2_{CF}}^S &= 58 \pm 2\,, &  c_{2_{CA}}^S & = - 60 \pm 1\,, & c_{2_{nf}}^S & = 43 \pm 1\,.
\end{align}
These constants were explored previously in \cite{Catani:1992ua}. 
Lacking the form of the
divergences near $\tau = 0$, these authors had to fit for the shape of the
curve as well as the constants, leading to results with much poorer accuracy. A comparison with the results of \cite{Catani:1992ua} is given Appendix \ref{sec:singular}.

\begin{figure}[t!]
\psfrag{N2}[]{\small $N^2$}
\psfrag{N0}[]{\small $N^0$}
\psfrag{Nm}[]{\small $1/N^{2}$}
\psfrag{Nnf}[]{\small $n_f N$}
\psfrag{NfNm}[]{\small $n_f/N$}
\psfrag{Nf2}[]{\small $n_f^{2}$}
\psfrag{sig}[t]{{\small $\phantom{aa}10^{-2}(1-T)$}{\Large $\frac{1}{\sigma}$}{\Large$\frac{\rd\sigma}{\rd T}$}}
\psfrag{log}[b]{\small $\phantom{abc}1-T$}
\begin{center}
\includegraphics[width=0.87\textwidth]{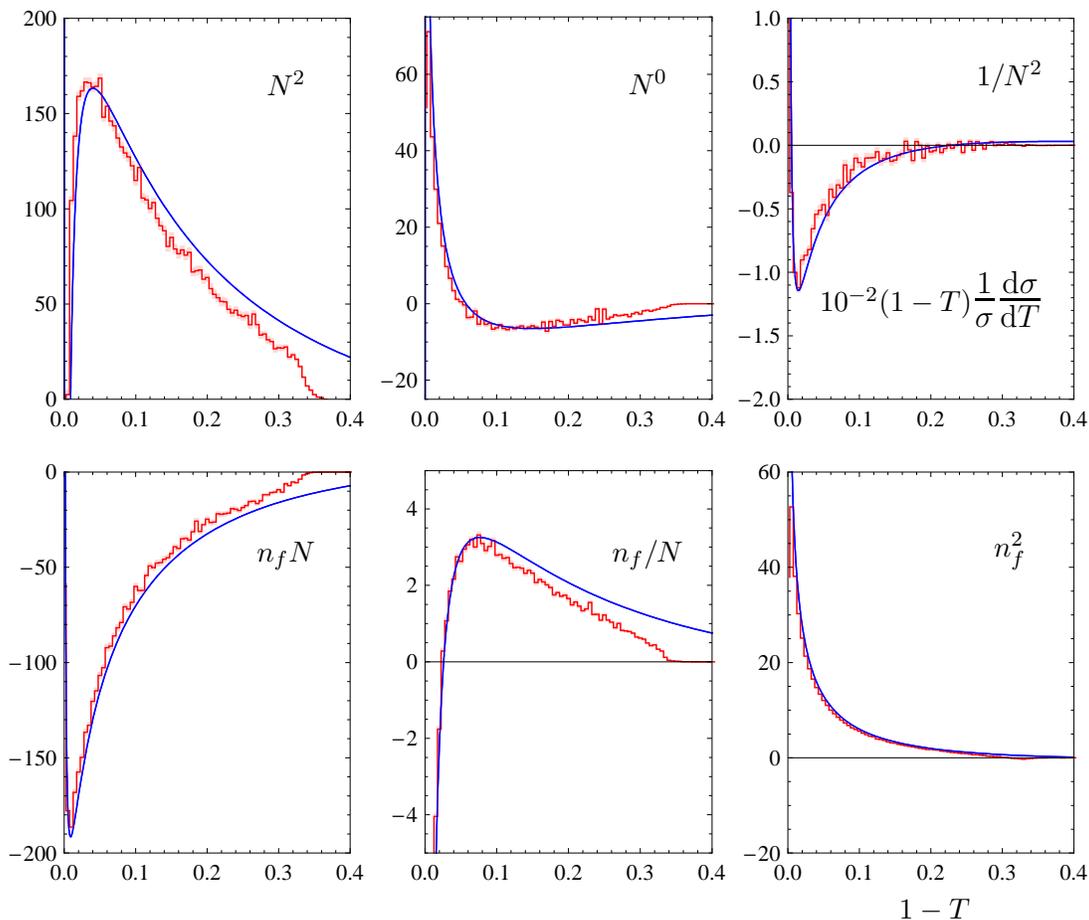} 
\end{center}\vspace*{-0.8cm}
\caption{\label{fig:NNLOcols} Contributions of different color structures to the three-loop coefficients of the thrust distribution. The plots show a comparison of our result for the singular terms encoded in $D_C$ (blue lines) with the numerical evaluation of the full coefficient $C$ (red histograms)  
\cite{GehrmannDeRidder:2007jk}. The light-red areas are an estimate of the statistical uncertainty.}
\end{figure}

\begin{figure}[t!]
\begin{center}
\psfrag{N2}[l]{\small $N^2$}
\psfrag{N0}[l]{\small $N^0$}
\psfrag{Nm}[l]{\small $1/N^{2}$}
\psfrag{Nnf}[l]{\small $n_f N$}
\psfrag{NfNm}[l]{\small $\phantom{a}n_f/N$}
\psfrag{Nf2}[l]{\small $n_f^{2}$}
\psfrag{sig}[B]{{\small $\phantom{ac}10^{-3}(1-T)$}{\Large $\frac{1}{\sigma}$}{\Large$\frac{\rd\sigma}{\rd T}$}}
\psfrag{log}[B]{\small $\phantom{ab}-\ln(1-T)$}
\includegraphics[width=0.90\textwidth]{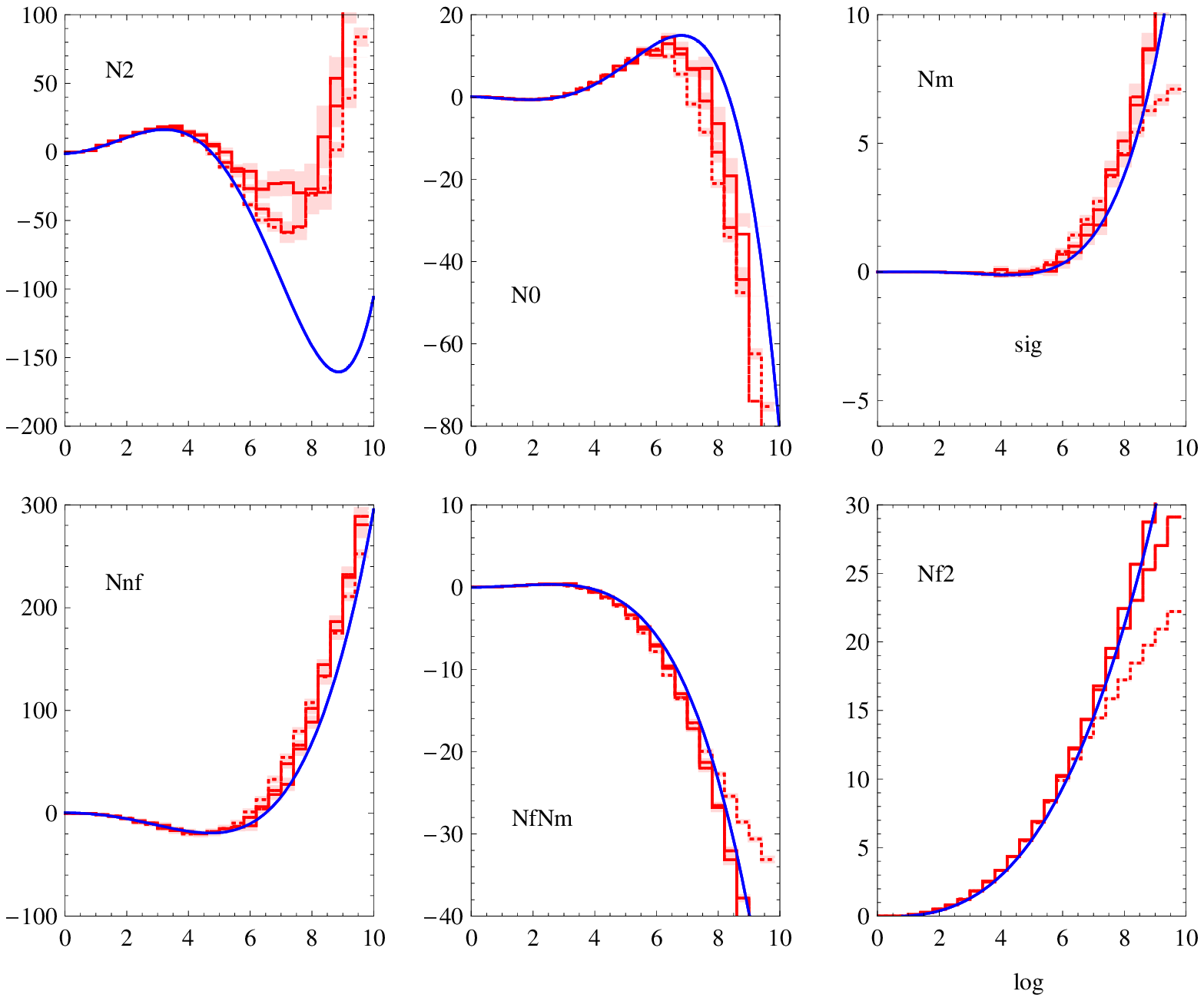}
\end{center}\vspace*{-0.8cm}
\caption{\label{fig:NNLOcolslog} Contributions to the three-loop coefficients of the thrust distribution. The plots show a comparison of our result for the singular terms (blue lines) with the numerical evaluation of the full result (red histograms)  \cite{GehrmannDeRidder:2007jk}. The dotted, dashed and solid lines correspond to an infrared cut-off $y_0=10^{-5}$,$10^{-6}$ and $10^{-7}$, see \cite{GehrmannDeRidder:2007jk}. The light-red areas are an estimate of the statistical uncertainty.}
\end{figure}

We now have all the necessary perturbative input at hand to evaluate the thrust distribution and to extract $\alpha_s$. Before doing so, we compare the recent NNLO fixed-order results in detail to the singular terms predicted using the effective theory. 
In Figure \ref{fig:NNLOcols} the contribution of the six color structures which appear at $\alpha_s^3$
to $C (\tau)$ and $D_C (\tau)$ are plotted.
 The color structure of the NNLO coefficient $C$ has the form $C = C_F (N^2 C_1 + C_2+ 1/N^{2} C_3 + N n_f C_4+ n_f/N C_5+ n_f^2 C_6 ) $ and the plot shows the six parts, with the prefactors evaluated for $N=3$  colors and $n_f=5$ quark flavors. The figure shows that the singular terms (blue lines) are a good approximation to the full result (red lines) for each color structure.  What is surprising is that they seem to agree well almost everywhere. One consequence of this is that the matching to the NNLO fixed-order distributions will have a small effect. The dominance of the logarithmically enhanced terms, even at moderate $\tau$, strongly suggests that resummation would indeed lead to a significant improvement in perturbative accuracy. The close agreement also provides a verification of the fixed-order result. 
Because the same numerical code is used for many other NNLO observables,
such an independent check is certainly welcome.

As we observed earlier, the lowest three bins of the NNLO fixed-order result of \cite{GehrmannDeRidder:2007jk} are higher than the singular terms obtained with the effective theory, see Figure $\ref{fig:ABC}$. The excess at small $\tau$ seen in Figure $\ref{fig:ABC}$ is barely noticeable in Figure \ref{fig:NNLOcols}, because we have multiplied the distributions by $\tau$  which de-emphasizes the small-$\tau$ region. To analyze this region in detail, we plot the distribution as a function of $\ln\tau$  in Figure \ref{fig:NNLOcolslog}. 
For very small $\tau$, the full result should reduce to the singular terms derived in the effective theory.
However, this region is very challenging for the numerical integration. The numerical results are shown in red in Figure \ref{fig:NNLOcolslog} and the light-red bands are the statistical uncertainty from the numerical NNLO calculation. The three red lines correspond to different values of an infrared cutoff, which is imposed when generating events \cite{GehrmannDeRidder:2007jk}. The agreement is good, except for the two leading color structures. 
The authors of~\cite{GehrmannDeRidder:2007jk} are aware of the problem  \cite{Gehrmann}.
For the extraction of $\alpha_s$, the region of very small $\tau$ will not be used, so these numerical difficulties
are not critical for present purposes.

\section{$\alpha_s$ extraction and error analysis \label{sec:fit}}

\begin{figure}[t]
 \begin{center}
 \includegraphics[width=\textwidth]{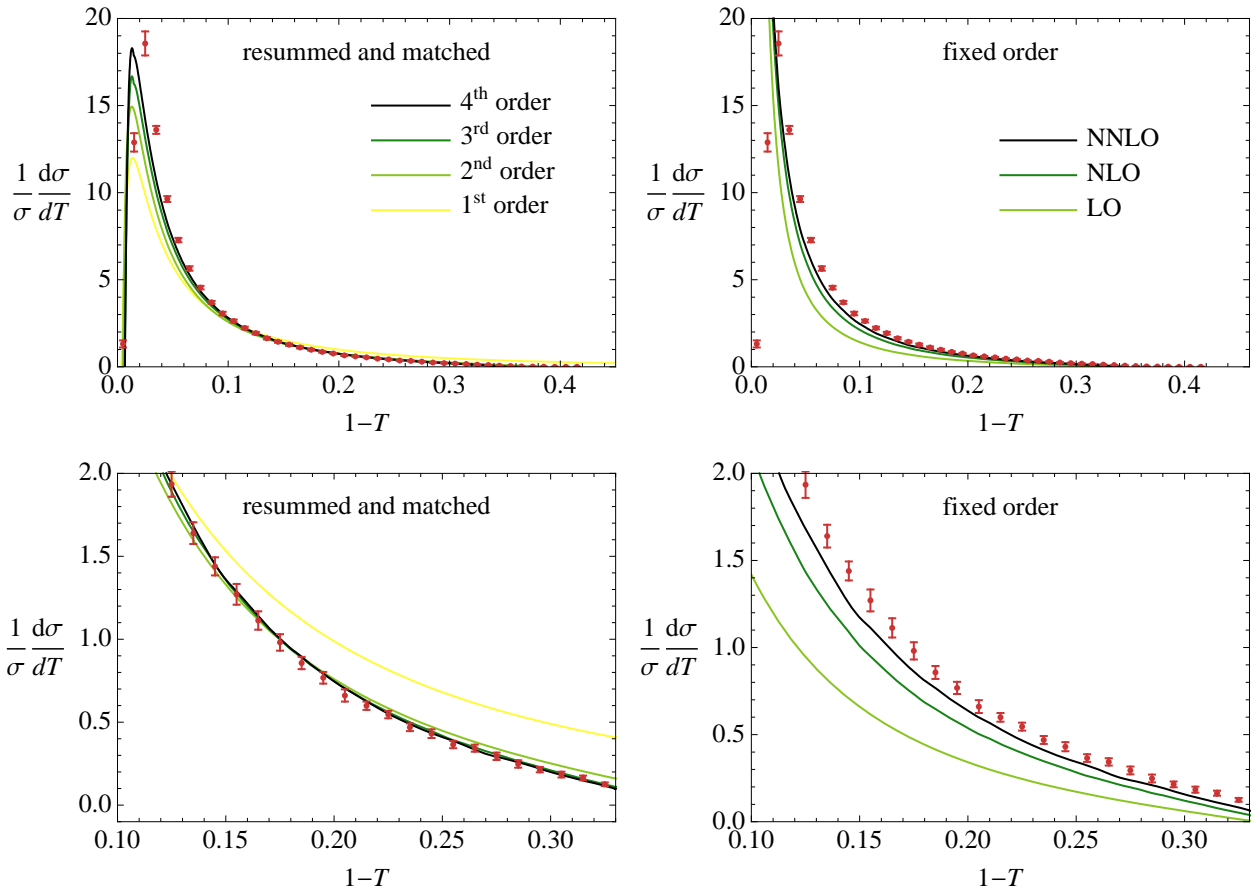}
 \end{center}
 \vspace{-0.8cm}
\caption{Convergence of resummed and fixed-order distributions. {\sc{aleph}} data (red) and {\sc{opal}} data (blue)
at $91.2$ GeV are included for reference. All plots have $\alpha_s(m_Z)=0.1168$.}
\label{fig:sfconv}
\end{figure}

In this section we now use our result for the thrust distribution to determine $\alpha_s$, using \lep data from {\sc{aleph}} \cite{Heister:2003aj} and {\sc{opal}} \cite{Abbiendi:2004qz}.  Before performing the fit, let us compare the perturbative expansion with and without resummation. The result at $Q=91.2$ GeV is shown in Figure \ref{fig:sfconv} side-by-side with the fixed-order expression. We use the same value $\alpha_s (m_Z) = 0.1168$ for both plots and have set the scales $\mu_h$, $\mu_j$ and $\mu_s$ to their canonical values (\ref{canonical}). For reference, we also show the {\sc{aleph}} and 
{\sc{opal}} data. The curves for the fixed-order calculation correspond to the standard LO, NLO, NNLO series; for the effective field theory calculation, the orders are defined in~Table~\ref{tab:ords}. 
It is quite striking how much faster the resummed distribution converges. 
In fact, it is hard to even distinguish the higher order curves after resummation, except in the region of very small $\tau$, where the distribution peaks. The peak region is affected by non-perturbative effects, as will be discussed in the next section, but it will not be used in the extraction of $\alpha_s$. 
The region relevant for the $\alpha_s$ extraction is shown in the lower two plots.
The value of  $\alpha_s (m_Z) = 0.1168$ we use in the plots corresponds to the best fit value in the range $0.1<\tau<0.24$ for the {\sc aleph} data set. However, the plot makes it evident that the extracted $\alpha_s$ value will not change much beyond first order. A fit to the NNLO fixed-order prediction gives $\alpha_s(m_Z) = 0.1275$. 

The {\sc{aleph}} and {\sc{opal}} collaborations have published 
analyses of the \lepone and higher energy \leptwo thrust
distributions.
 To fit $\alpha_s$ we calculate the thrust distribution integrated over each
bin measured in the experiments. The resummed contribution in a given bin is obtained as $R_2(\tau_R) - R_2(\tau_L)$ using Eq.~\eqref{R2eq}  for the bin with $\tau_L < \tau < \tau_R$. For the matching contribution, we integrate analytically the $D_A(\tau), D_B(\tau)$ and $D_C(\tau)$ functions and subtract them from the analytic integral of $A (\tau)$ and the appropriately binned numerical distributions $B (\tau)$ and $C (\tau)$. 

 A problem we encounter when trying to extract $\alpha_s$ is that the experiments have published statistical, systematic, and hadronization uncertainties for each bin, but have not made the bin-by-bin correlations public. Without this information, we proceed with a conservative approach to error estimates: to extract the default value of $\alpha_s$, we perform a $\chi^2$-fit to the data including only statistical uncertainties. We then use the systematic and hadronization errors on $\alpha_s$ obtained in previous fits to {\sc{aleph}}  \cite{Dissertori:2007xa} and {\sc{opal}} \cite{Abbiendi:2004qz} data. In these papers fits to $\alpha_s$ were performed which included the correlation information. To be able to use their values, we perform our fits using exactly the same fit ranges as used in these papers. This is not entirely optimal, since the experimental systematic error will depend somewhat on the theoretical model used in the fit.
\begin{figure}[t]
\psfrag{x}[l]{$\!\!\!\!\!\frac{\mathrm{total\: exp.\: uncertainty}}{\mathrm{data}}$}
\psfrag{y}[l]{$\!\!\!\!\!\frac{\mathrm{stat.\:\: uncertainty}}{\mathrm{data}}$}
\psfrag{z}[l]{$\!\!\!\!\!\frac{\mathrm{fit-data}}{\mathrm{data}}$}
\begin{center} 
\includegraphics[width=0.7\textwidth]{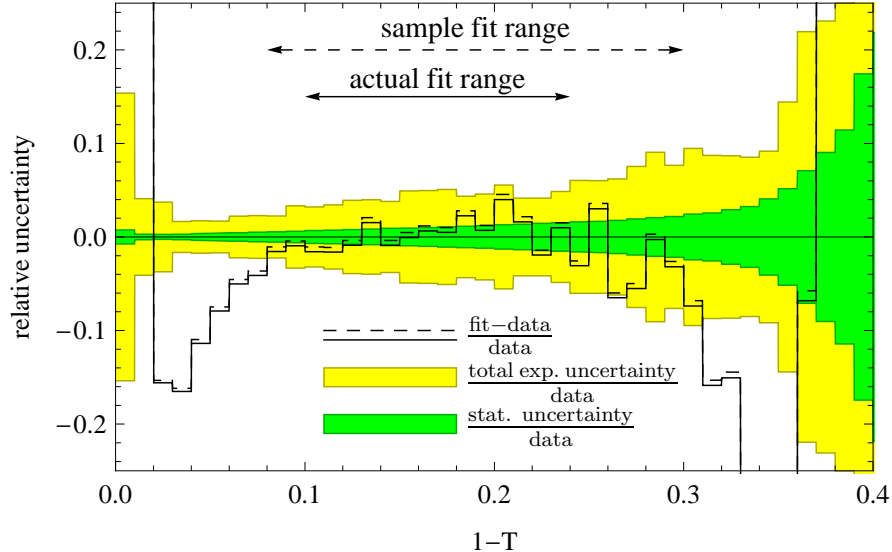}
\end{center}
\vspace*{-0.8cm}
\caption{Relative error for best fit to {\sc aleph} data at $91.2$ GeV. 
The inner green band includes only
statistical uncertainty, while the outer yellow band includes statistical, systematic and hadronization
uncertainties. The solid line is fit to $0.1 < 1-T < 0.24$ giving $\alpha_s(m_Z)=0.1168$ while
the dashed line is fit from $0.08 < 1-T < 0.3$ giving $\alpha_s(m_Z) = 0.1171$. The smaller
fit range is used for the error analysis because it has been previously studied
in~\cite{Dissertori:2007xa}.}
\label{fig:chis}
\end{figure}
Our resummed calculation is valid in a wider range of $\tau$ than the predictions used in \cite{Dissertori:2007xa, Abbiendi:2004qz}, so one could use data closer to the peak, where the statistics are higher and resummation is more important. In a future analysis, the fit range could be optimized to minimize the total error after folding in the proper correlations.

 In Figure \ref{fig:chis}, we plot the relative statistical and total experimental uncertainty as a function of $\tau$ and compare to the best fit result. We find that the extracted value is fairly insensitive to the fit range. In fact, going from the standard range (solid line) to the larger region (dashed lines) changes the best-fit value of $\alpha_s (m_Z)$ by less than 0.3\%, from 0.1168
to 0.1171. 

\begin{figure}[t!]
\psfrag{y}[]{{\scriptsize $(1-T)$} {\small $\frac{1}{\sigma}\frac{\rd \sigma}{\rd T}$}}
\psfrag{x}[]{\scriptsize $1-T$}
\begin{center}
\begin{tabular}{ll}
\includegraphics[width=0.47\textwidth]{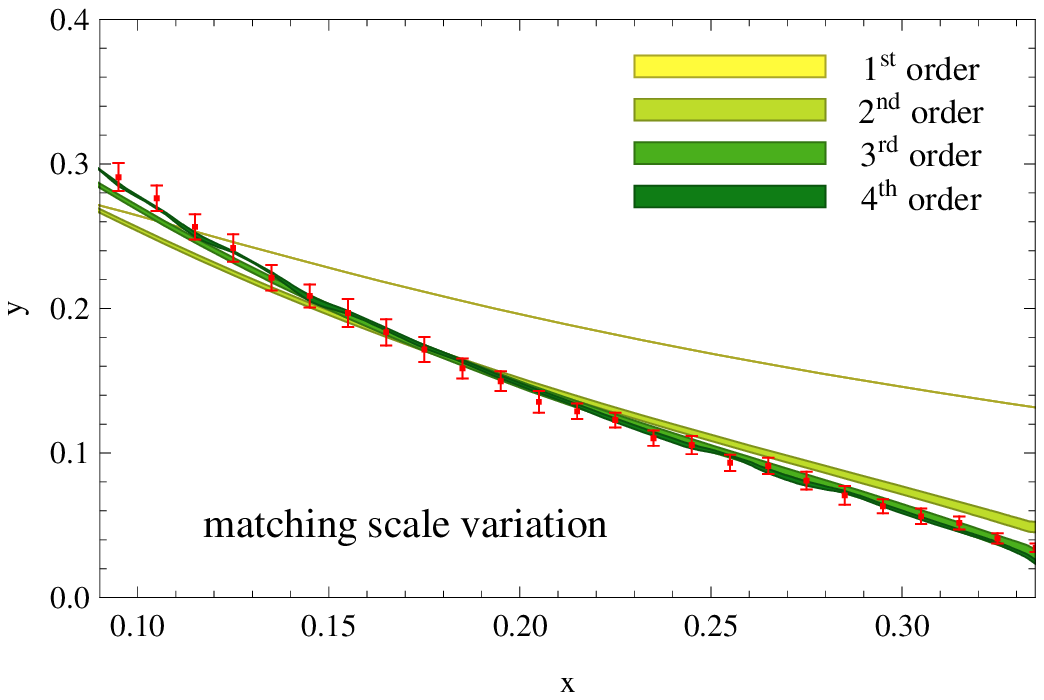} & \includegraphics[width=0.47\textwidth]{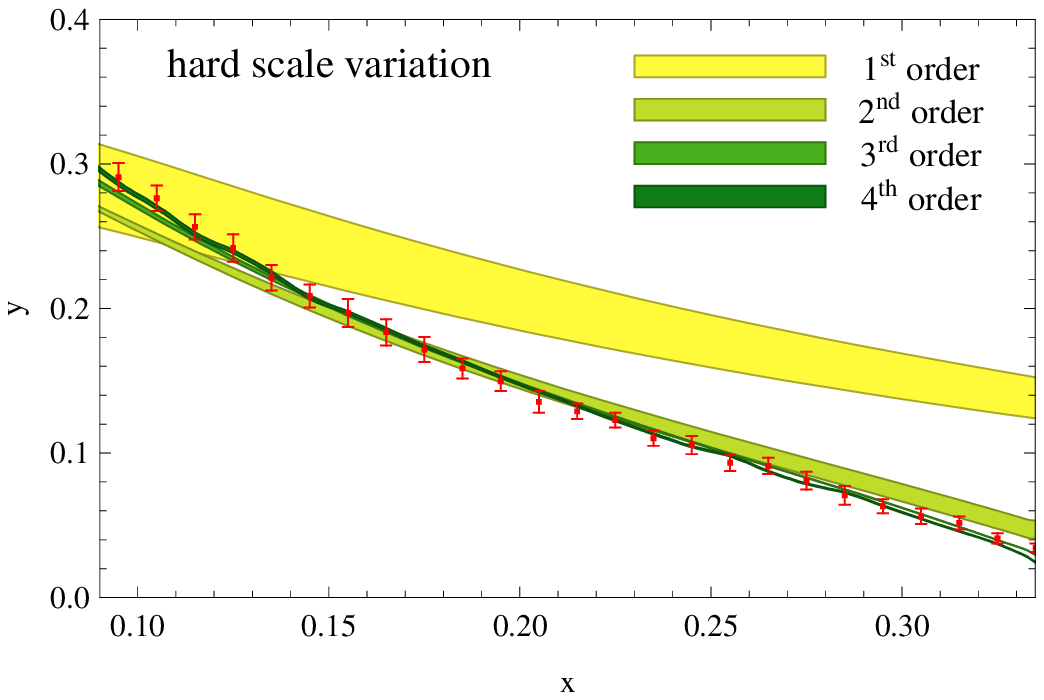} \\[6pt]
\includegraphics[width=0.47\textwidth]{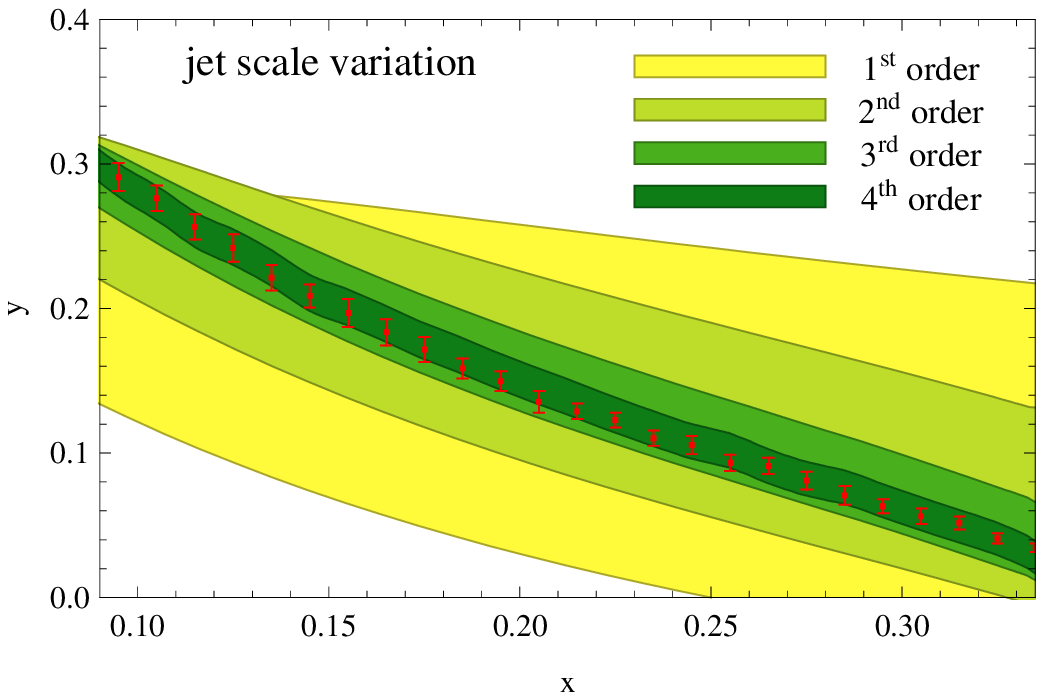} & \includegraphics[width=0.47\textwidth]{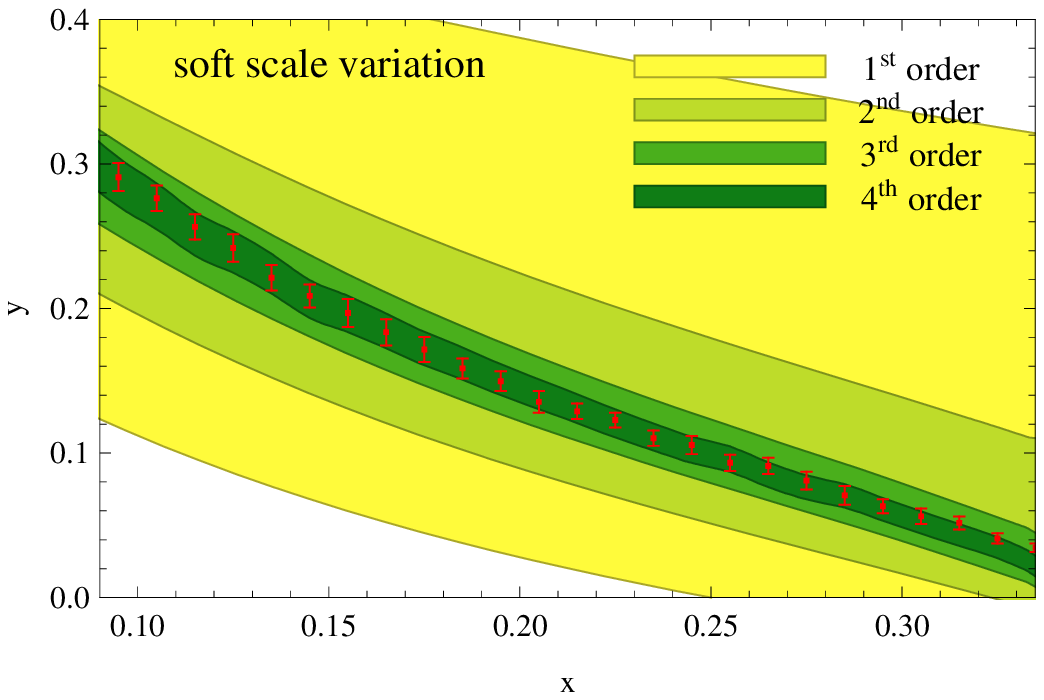}  \\[6pt]
\includegraphics[width=0.47\textwidth]{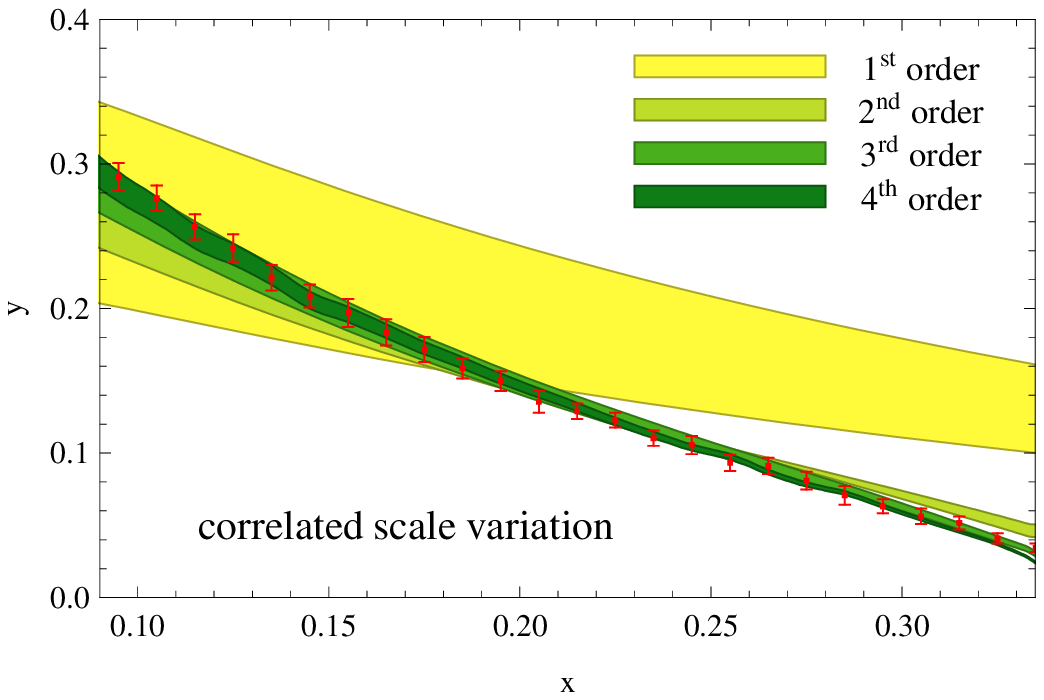} & \includegraphics[width=0.47\textwidth]{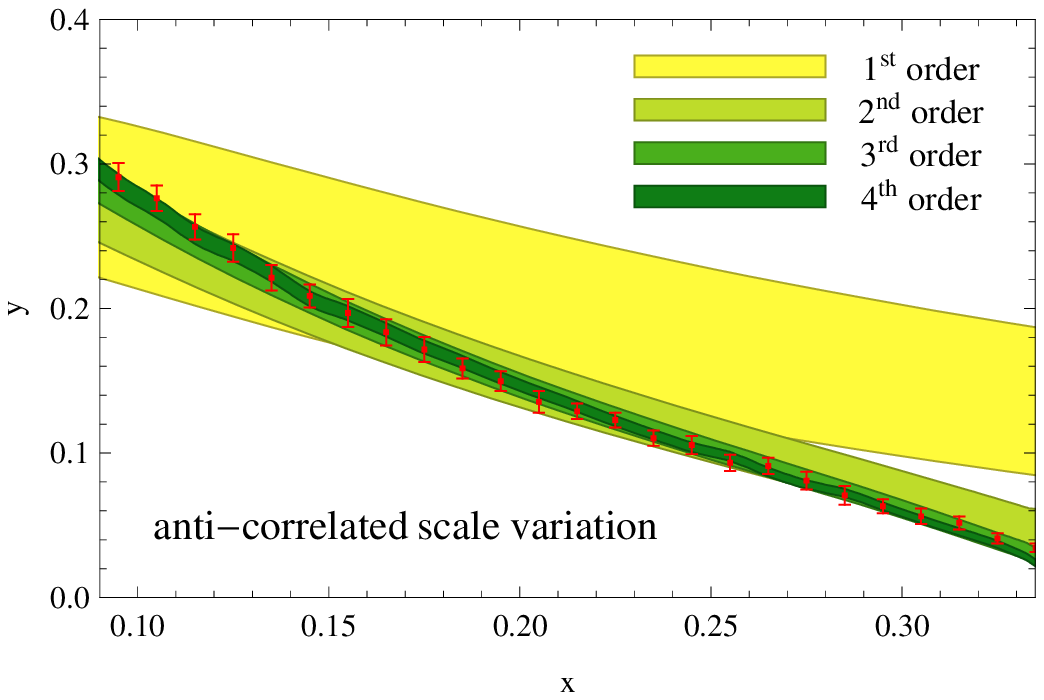}
\end{tabular}
\end{center}
\vspace{-0.6cm}
\caption{Perturbative uncertainty at $Q=91.2\,{\rm GeV}$. The first four panels show the variation of the matching scale, the hard scale, the jet scale, and the soft scale. 
Each of the scales is varied separately by a factor of two around the default value. The last two panels show the effect of simultaneously varying the jet- and soft scales, see text. The \lepone  {\sc aleph} data is included for reference. All plots have $\alpha_s(m_Z)=0.1168$.}
\label{fig:MHJSBC}
\end{figure}

Next, we consider the perturbative theoretical uncertainty. In the effective
field theory analysis, four scales appear: the hard scale $\mu_h \sim Q$, the jet
scale $\mu_j \sim \sqrt{\tau} Q$, the soft scale $\mu_s \sim \tau Q$, and the
scale $\mu_m$ at which the matching corrections are added. In the matching corrections the physics associated with the hard, jet and soft scales has not been factorized, so it is not obvious which value of $\mu_m$ should be chosen. We follow standard fixed-order practice and choose $\mu_m=Q$ as the default value. Our result is independent of these scales to the order of the calculation: the change in the result due to scale variation can thus be used to estimate the size of unknown higher order terms, of $O(\alpha_s^4)$ for our final result. 

We show the
results of varying each of the four scales up and down by a factor of 2 in the first four panels of Figure
\ref{fig:MHJSBC}. The results converge nicely, with the dominant uncertainty coming from
the soft scale variation. This is expected, as the soft scale probes the
lowest energies and therefore the largest values of $\alpha_s$. In fact, it is
a critical advantage of the effective theory that the soft scale can be probed
explicitly -- the fixed-order calculation has access to only one scale and assuming $\mu\sim Q$ may therefore underestimate the perturbative uncertainty.
From the first panel in Figure \ref{fig:MHJSBC} it is clear that the extraction of
$\alpha_s$ is almost completely insensitive to the scale at which the fixed
order calculation comes in. Again, this is in contrast to a pure fixed-order
result. The matching scale variation is 
so small because the matching correction itself is small, as we saw in 
Figures \ref{fig:ABC}, \ref{fig:NLOcols}, \ref{fig:NNLOcols}, and \ref{fig:NNLOcolslog}.

Figure \ref{fig:MHJSBC} shows the effect of varying the jet and soft scales separately by factors of two:
$\frac{1}{2} \sqrt{\tau} Q < \mu_j < 2 \sqrt{\tau} Q$ and $\frac{1}{2} \tau Q
< \mu_s < 2 \tau Q$. While a factor of two may seem reasonable for a fixed
order calculation (although as we have already observed, the thrust
distribution probes scales $\tau Q \ll Q$), from the effective
field theory point of view it makes little sense  to vary the soft and jet scales separately. In doing so, one can easily have $\mu_j < \mu_s$ or  $\mu_h < \mu_j$ which
is completely unphysical. Instead, for the error analysis we will use two
coordinated variations. First, a correlated variation holding $\mu_j / \mu_s$ fixed:
\begin{equation}
  \mu_j \rightarrow c \sqrt{\tau} Q, \hspace{1em} \mu_s \rightarrow c \tau Q,
  \hspace{1em} \text{$\frac{1}{2} < c < 2$}\,.
\end{equation}
This probes the upper and lower limits on $\mu_j$ and $\mu_s$, but avoids the
unphysical region. Second, an anti-correlated variation, holding
$\mu_j^2 / (Q \mu_s)$ fixed:
\begin{equation}
  \mu_j^2 \rightarrow a Q^2  \tau \hspace{1em} \mu_s \rightarrow a Q  \tau,
  \hspace{1em} \frac{1}{\sqrt{2}} < a < \sqrt{2}\,.
\end{equation}
This is independent from the correlated mode but again avoids
having $\mu_j < \mu_s$. 
The uncertainty resulting from these two variations is shown in the last two panels of Figure \ref{fig:MHJSBC}.

\begin{figure}
  \begin{minipage}[]{.9\textwidth}
    \begin{center}  
      \epsfig{file=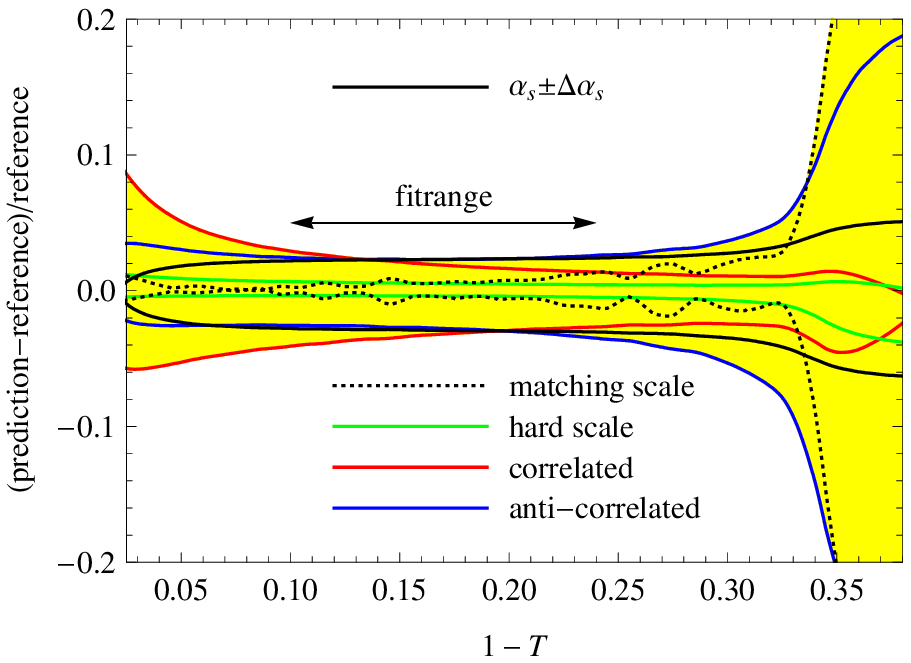, scale=1.3}
    \end{center}
  \end{minipage}
  \begin{minipage}[]{.9\textwidth}
    \begin{center}  
      \epsfig{file=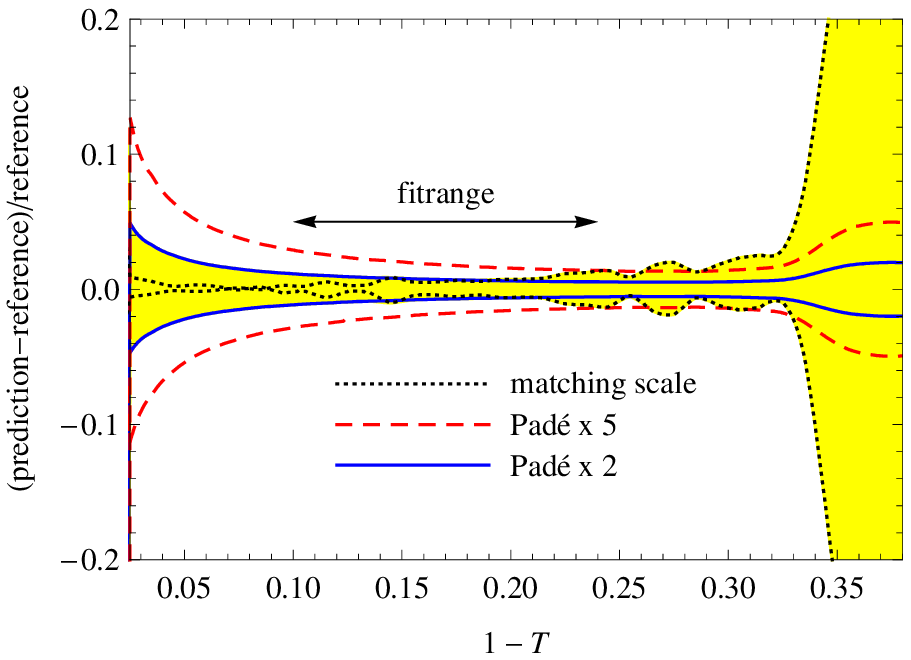,scale =1.3}
    \end{center}
  \end{minipage}
\caption{Uncertainty bands for various scale variations. The band in the first panel
is determined entirely by scale variations.
The second panel
shows an alternative way of estimating the perturbative uncertainty using an educated guess
of the uncalculated higher order coefficients, as described in the text. 
}
\label{fig:uncbands}
\end{figure}

To estimate the total  perturbative uncertainty on the extracted value of $\alpha_s$, we use the uncertainty band technique proposed in \cite{Jones:2003yv} and adopted both by {\sc{aleph}} \cite{Heister:2003aj} and {\sc{opal}} \cite{Abbiendi:2004qz} as well as in the recent fit of NNLO results to {\sc{aleph}} data \cite{Dissertori:2007xa}. 
The result is shown in Figure~\ref{fig:uncbands}.
In short, the theoretical uncertainty is determined as follows: one first calculates $\alpha_s(m_Z)$ using a least-squares fit to the data
with all scales at their canonical values and without including any theoretical uncertainty in the $\chi^2$-function. 
Then each scale is varied
separately, holding $\alpha_s (m_Z)$ fixed to its best-fit value. These produce the curves in
Figure \ref{fig:uncbands}. Next, the uncertainty band, the yellow region in 
Figure \ref{fig:uncbands}, is defined as the envelope of  all these variations. 
Finally, the scales are returned to their canonical values, 
and the maximal and minimal values of $\alpha_s$ are determined which allow the prediction 
to remain within the uncertainty band. 
An important feature of this approach is that the data enters only in the determination of
 the best fit $\alpha_s$ and the fit region; 
the perturbative uncertainty is determined purely from within the theoretical calculation.
Separating the theoretical and experimental errors in this way makes 
it much easier to average $\alpha_s$ results obtained from different data sets, 
since they suffer from the same theoretical uncertainty.

\begin{table}[t!]
\begin{center}
  \begin{tabular}{|c|c|c|c|c|c|c|c|c|c|} 
    \hline
    Q    & 91.2 & 133  & 161  &  172 &  183 & 189  & 200  & 206  & AVG\\ \hline  \hline
\multirow{2}{*}{fit range}   &  0.1  & 0.06 & 0.06& 0.04 & 0.06 & 0.04 & 0.04 & 0.04 & \multirow{2}{*}{--}\\
  & 0.24  & 0.25 & 0.25 & 0.2 & 0.25 &0.2 & 0.2 & 0.2 & \\ 
\hline
       $\chi^2$/d.o.f.\ & 32.5/13 & 7.7/4 & 3.3/4 & 10.3/4 & 3.6/4 & 0.9/4 & 24.6/4 & 4.0/4 & -- \\
stat.\ err.\ & 0.0001 & 0.0037 & 0.0070 & 0.0080 & 0.0043 & 0.0022 & 0.0023 & 0.0023 & 0.0010 \\
 syst.\ err.\ & 0.0008 & 0.0010 & 0.0010 & 0.0010 & 0.0011 & 0.0010 & 0.0010 & 0.0010 & 0.0010\\
 hadr.\ err.\ & 0.0019 & 0.0014 & 0.0012 & 0.0012 & 0.0011 & 0.0011 & 0.0010 & 0.0010 & 0.0012\\
 pert.\ err.\ & $^{+0.0013}_{-0.0017}$ & $^{+0.0012}_{-0.0016}$ & $^{+0.0015}_{-0.0020}$ &
   $^{+0.0006}_{-0.0009}$ & $^{+0.0010}_{-0.0013}$ & $^{+0.0011}_{-0.0015}$ & $^{+0.0010}_{-0.0014}$ &
   $^{+0.0009}_{-0.0012}$ & 0.0012 \\
tot.\ err.\  & 0.0026 & 0.0043 & 0.0074 & 0.0082 & 0.0047 & 0.0030 & 0.0030 & 0.0029 & 0.0022 \\
\hline
(Pad\'e $\times$ 2) & 0.0004 & 0.0004 & 0.0003 & 0.0004 & 0.0004 & 0.0004 & 0.0004 & 0.0003 & -- \\
\hline
 {$\alpha_s(m_Z)$} & 0.1168 & 0.1183 & 0.1263 & 0.1059 & 0.1160 & 0.1203 & 0.1175 & 0.1140 &0.1168 \\  
\hline
\hline
{\sc pythia} & 0.1152 & 0.1164 & 0.1248 & 0.1028 & 0.1146 & 0.1177 & 0.1151 & 0.1119 & 0.1146 \\
{\sc ariadne} & 0.1169 & 0.1181 & 0.1264 & 0.1047 & 0.1164 & 0.1197 & 0.1170 & 0.1135 & 0.1164 \\
\hline
  \end{tabular}
  \caption{Best fit to {\sc aleph} data. \label{tab:alephresults} The row labeled (Pad\'e $\times$ 2) is
an alternative measure of perturbative uncertainty as described in the text. It is not combined into the total error.
The rows labeled {\sc pythia}  and {\sc ariadne}  give 
the value of $\alpha_s$ after correcting for hadronization
and quark masses using {\sc pythia} or {\sc ariadne}.}
\end{center}
\end{table}

The purpose of scale variations is to estimate the effect that a higher order perturbative calculation would have
on a distribution. This is justified by arguing that any scale variation can be compensated by terms at one order
higher in $\alpha_s$, thus it should give a reasonable estimate of these higher order terms. However, as we have seen, 
the amount by which we vary the scales is arbitrary, and the traditional factor of 2 in the variation is
both problematic for the jet and soft scales and seems to overestimate the uncertainty. The distribution
determined by the effective theory at one higher order depends on only a handful of numbers: the beta-function
coefficient $\beta_4$, the anomalous
dimensions $\Gamma_4, \gamma^H_3, \gamma^J_3$ and the constants in the hard, jet and soft functions,
$c^H_3, c^J_3, c^S_3$, see Appendix \ref{sec:anomalous} for the subscript conventions. Thus in the effective field theory, there is a  straightforward way
to estimate the effect of higher orders: one simply varies these coefficients. For example, we can estimate their size using a Pad\'e approximation: $\Gamma_{n+1}=\pm c \frac{\Gamma_n^2}{\Gamma_{n-1}}$. This
should reasonably span likely values for what a higher order perturbative calculation would provide. We show
the variations corresponding to $c=2$ and $c=5$ in the bottom panel of Figure~\ref{fig:uncbands}, which
are labeled Pad\'e $\times$ 2 and Pad\'e $\times$ 5 respectively. 
In each case we scan over the signs for the various coefficients
to find the largest variations. Even for $c=2$, the fifth order coefficients come out quite large, for
example, $\Gamma_4 \approx \pm 2 \times 10^4$ and $\beta_4 \approx \pm 3\times 10^5$. Nevertheless, the uncertainty is
still significantly smaller than what we found using scale variation. We find that Pad\'e $\times$ 2 gives
$\delta \alpha_s(m_Z) \sim 0.0003$ in contrast to errors around $\delta \alpha_s(m_Z)\sim 0.0012$
from the scale variations. Although the higher order constants are unknown, one might try to
estimate them in more sophisticated ways, for example, by computing the dominant diagrams. 
In the end, we will not use this new method
for the final error estimates, but we present the resulting uncertainties in 
Tables \ref{tab:alephresults} and \ref{tab:opalresults} for completeness. They
 include a scale
variation in the matching correction because this is independent of the resummed distribution. 

For each of the energies in the {\sc{aleph}} and {\sc{opal}} data
 sets, we perform a least-squares fit using the experimental statistical errors. 
The statistical uncertainty on $\alpha_s$
is calculated from the variation in $\chi^2$, the perturbative uncertainty is
calculated using the uncertainty band (with scale variations), and systematic uncertainties are taken
from~\cite{Dissertori:2007xa} and~\cite{Abbiendi:2004qz}, as discussed above. 
We include the non-perturbative hadronization uncertainties from these papers, but do not include the corresponding hadronization corrections. We will discuss hadronization and other power-suppressed effects in detail in the next section. The fit results are given in Tables~\ref{tab:alephresults}~and~\ref{tab:opalresults}.

\begin{table}[t!]
\begin{center}
  \begin{tabular}{|c|c|c|c|c|c|}
    \hline
    Q & 91.2 & 133 & 177 & 197 & AVG\\
    \hline
    \hline
    fit range & 0.05-0.3 & 0.05-0.3 & 0.05-0.3 & 0.05-0.3 & --\\
\hline
    $\chi^2$/d.o.f. &  149.9/5 & 17.0/5 & 1.7/5 & 18.3/5 & -- \\
stat.\ err.& 0.0001 & 0.0038 & 0.0033 & 0.0014 & 0.0014\\
syst.\ err. & 0.0011 & 0.0054 & 0.0028 & 0.0013 &  0.0013 \\
hadr.\ err. & 0.0031 & 0.0024 & 0.0021 & 0.0019 & 0.0019  \\
 pert.\ err. &  $^{+0.0014}_{-0.0018}$ &  $^{+0.0011}_{-0.0015}$ &  $^{+0.0009}_{-0.0013}$ &
  $^{+0.0011}_{-0.0014}$ & 0.0013 \\
 tot.\ err. & 0.0037 & 0.0072 & 0.0049 & 0.0030 & 0.0030\\
\hline
(Pad\'e $\times$ 2) & 0.0004 & 0.0003 & 0.0003 & 0.0003 &  -- \\
\hline
  $\alpha_s(m_Z)$ & 0.1189 & 0.1165 & 0.1153 & 0.1189 &  0.1189  \\
\hline
\hline
{\sc pythia}& 0.1143 & 0.1142 & 0.1134 & 0.1173 &  0.1173  \\
{\sc ariadne} & 0.1163 & 0.1160 & 0.1151 & 0.1189 &   0.1189  \\
\hline
  \end{tabular}
  \caption{Best fit to {\sc opal} data. \label{tab:opalresults}}
\end{center}
\end{table}

To combine the results from different energies, we compute a weighted average, 
$\bar\alpha_s = \sum_i w_i \alpha_s^{(i)}$. The weights $w_i$ are determined by minimizing the uncertainty $\bar \sigma^2 =\sum_{ij} w_i\, w_j \,{\rm cov}(i,j)$. Given that we don't know the exact correlations, we set
\begin{equation}
{\rm cov}(i,j) = \left(\sigma_{\rm stat}^{(i)}\right)^2 \delta_{i,j} +  \sigma_{\rm sys}^{(i)} \sigma_{\rm sys}^{(j)} 
+  \sigma_{\rm hadr}^{(i)} \sigma_{\rm hadr}^{(j)} +\sigma_{\rm pert}^{(i)} \sigma_{\rm pert}^{(j)}  \,.
\end{equation}
That is, we assume uncorrelated statistical errors and 100\% correlation for the systematic, hadronic and perturbative uncertainties at different energies. Because the correlated uncertainties are dominant, naively minimizing the uncertainty can in some cases be lead to negative weights. This happens when combining the OPAL results in the above way. We eliminate these solutions by imposing $w_i>0$, after which the best value from OPAL is obtained by assigning 100\% weight to the highest energy measurement which has the smallest systematic uncertainty. The result obtained  after combining {\sc{aleph}} and {\sc{opal}} results individually is given in the last column in Tables \ref{tab:alephresults} and \ref{tab:opalresults}. Finally, we combine the {\sc{aleph}} and  {\sc{opal}} results to an overall average. In this case, we assume that the systematic uncertainties are completely correlated between the individual energy results from each experiment, but neglect the correlations between the systematical uncertainties among the two experiments. For the hadronization and perturbative error, we assume 100\% correlation. Proceeding in this way, we find
 \begin{align}
  \alpha_s (m_Z) &= 0.1172 \pm 0.0010 (\mathrm{stat}) \pm 0.0008 (\mathrm{sys}) \pm
  0.0012 (\mathrm{had}) \pm 0.0012 (\mathrm{pert}) \nonumber \\
& = 0.1172 \pm 0.0022\,.
\end{align}
This result is close to the PDG world average $\alpha_s (m_Z) = 0.1176
\pm 0.0020$ and has similar uncertainties. 

Our calculation does not include hadronization corrections and neglects quark masses. If we estimate their effect using {\sc{pythia}}, the central value shifts to $\alpha_s (m_Z) = 0.1150$, while correcting with {\sc{ariadne}} gives $\alpha_s (m_Z) = 0.1168$. We observe that the difference we find between {\sc{pythia}} and {\sc{ariadne}} is larger than the hadronization uncertainty in our average, which is based on {\sc{aleph}} and {\sc{opal}} studies.
Correcting our higher order perturbative result with a tuned leading-order Monte Carlo shower is problematic, so this difference should be interpreted with caution. Various issues associated with hadronization corrections will be discussed in detail in the next section.  
 
 \begin{figure}[t!]
\begin{center}
\begin{tabular}{cc}
\includegraphics[width=0.48\textwidth]{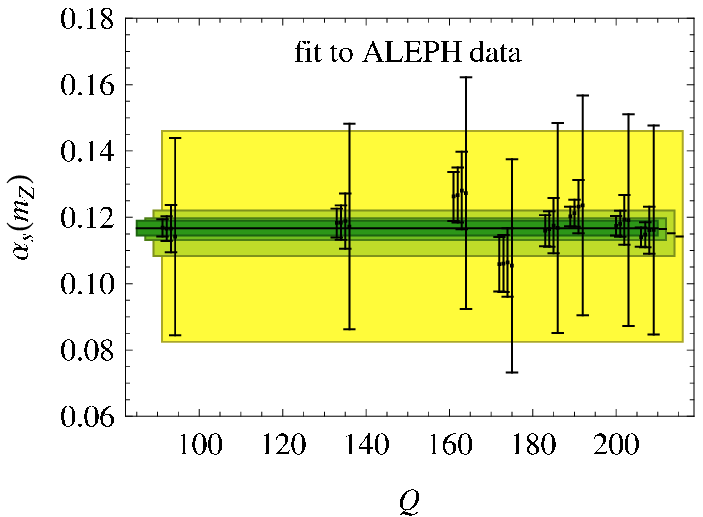} &
\includegraphics[width=0.48\textwidth]{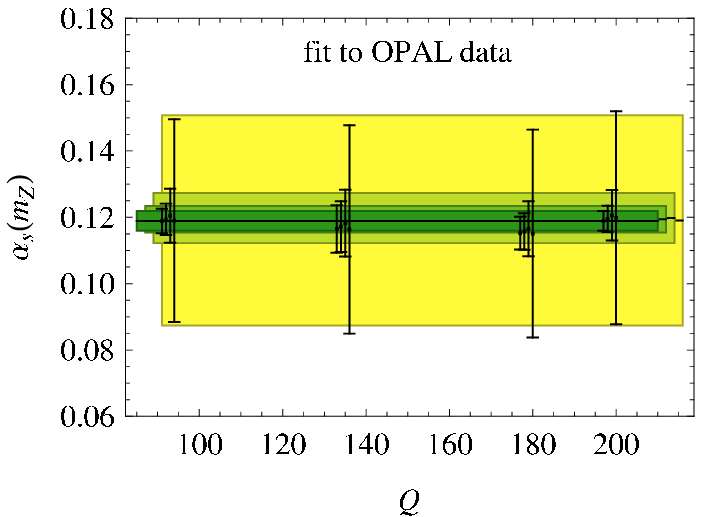}
\end{tabular}
\end{center}
\vspace{-0.8cm}
\caption{Best fit values for $\alpha_s(m_Z)$. From right to left the lines are the total error bars at each energy for
 first order, second order, third order and fourth order, as defined in the text. The bands are weighted averages
with errors combined from all energies.}
\label{fig:fitresults}
\end{figure}

It is interesting to repeat the fit order by order.
This is done in Table \ref{tab:fitorder} and displayed graphically in Figure~\ref{fig:fitresults}. 
The figure shows that the results found at different energies are consistent and illustrates the reduction of the uncertainty when including higher order terms.

\begin{table}[t!]
\begin{center}
 \begin{tabular}{|c|c|c|c||c|c|c|}  \hline
 \multicolumn{7}{|c|} {\sc{aleph}}  \\ \hline
 &  \multicolumn{3}{|c||}{\lepone +\leptwo} & \multicolumn{3}{c|}{\lepone}  \\ \hline
    order  & $\alpha_s$ & total err  & pert. err & $\alpha_s$  & tot.err & pert.err  \\  \hline
    \first order& 0.1142 & 0.0297 & 0.0296 & 0.1142 & 0.0297 & 0.0296 \\
\second order& 0.1152 & 0.0068 & 0.0064 & 0.1166 & 0.0071 & 0.0068 \\
 \third order& 0.1164 & 0.0033 & 0.0027 & 0.1166 & 0.0037 & 0.0031 \\
 \fourth order& 0.1168 & 0.0022 & 0.0012 & 0.1168 & 0.0026 & 0.0015 \\ \hline
\multicolumn{7}{c}{} \\  \hline
 \multicolumn{7}{|c|} {\sc{opal}} \\  \hline
&  \multicolumn{3}{|c||}{\lepone + \leptwo} & \multicolumn{3}{c|}{\lepone}  \\ \hline
    order  & $\alpha_s$ & total err  & pert. err & $\alpha_s$  & tot.err & pert.err  \\  \hline
    \first order& 0.1190 & 0.0305 & 0.0304 & 0.1190 & 0.0305 & 0.0304 \\
  \second order & 0.1198 & 0.0076 & 0.0070 & 0.1205 & 0.0081 & 0.0074 \\
 \third order& 0.1194 & 0.0040 & 0.0029 & 0.1194 & 0.0047 & 0.0034 \\
 \fourth order& 0.1189 & 0.0030 & 0.0013 & 0.1189 & 0.0037 & 0.0016\\ 
 \hline
  \end{tabular}
  \caption{Best fit values and uncertainties at different orders, as defined in Table \ref{tab:ords}.}
  \label{tab:fitorder}
\end{center}
\end{table}

\section{Non-perturbative effects and quark mass corrections \label{sec:NP}}

Let us now turn to two power suppressed effects which we have so far neglected in our analysis. The first is hadronization: the effective theory calculation corresponds to a parton-level distribution, while the experiment measures hadrons. Secondly, we have neglected quark masses in our calculation. Because thrust is an infrared-safe observable, both corrections are expected to be small, however they may not be negligible.

Most of the previous determinations of $\alpha_s$ have used Monte Carlo event generators to correct the parton-level predictions for hadronization effects and estimate the hadronic uncertainty by comparing the output of different generators. In particular, {\sc{aleph}} \cite{Heister:2003aj},  {\sc{opal}} \cite{Abbiendi:2004qz} and the recent NNLO analysis \cite{Dissertori:2007xa} all use {\sc{pythia}} to obtain their default hadronization corrections and then compare to {\sc{herwig}} and {\sc{ariadne}} to obtain the associated uncertainty. It turns out that the largest differences generally occur between {\sc pythia} and {\sc ariadne} \cite{Abbiendi:2004qz}, even though {\sc ariadne} uses {\sc pythia} to calculate hadronization.

We include in Tables~\ref{tab:alephresults}~and~\ref{tab:opalresults} the best fit values of $\alpha_s$ obtained after correcting the data bin-by-bin for hadronization and $b$- and $c$-quark masses using {\sc pythia} {\tt v.6.409}~\cite{Sjostrand:2006za}, with default parameters, and with {\sc ariadne} {\tt v.4.12}~\cite{Lonnblad:1992tz}, using the {\sc aleph} tune. Correcting with {\sc ariadne} has quite a small effect on the values of $\alpha_s$.  Moreover, {\sc ariadne} always gives a larger value of $\alpha_s$ than {\sc pythia}. If the central values are taken after the {\sc ariadne} corrections, they agree quite closely with a fit to the parton level distributions, that is, without any hadronization. In addition, we also used the new {\sc sherpa} dipole shower \cite{Winter:2007ye} for hadronization and find results similar to {\sc ariadne}.

Relying on the Monte-Carlo generators for hadronization is clearly not ideal, since they have been tuned to the same \lep data we are trying to reproduce! The situation is especially problematic when trying to correct our resummed distribution. The Monte Carlo generators are all based on the parton-shower approximation, which only sums the leading Sudakov double logarithms and part of the next-to-leading logarithms.
In contrast, our distribution is correct to N$^3$LL and to NNLO in fixed-order perturbation theory. By tuning to data, part of the missing higher order perturbative corrections get absorbed into the hadronization model. An obvious way to avoid this problem would be to include the higher order corrections into the Monte Carlo codes, but needless to say, no such generator yet exists (although, see~\cite{Bauer:2006mk,Bauer:2006qp} for an approach to improving generators based on the same effective field theory ideas we are using).

\begin{figure}[t!]
\begin{center}
\begin{tabular}{cc}
\includegraphics[width=0.48\textwidth]{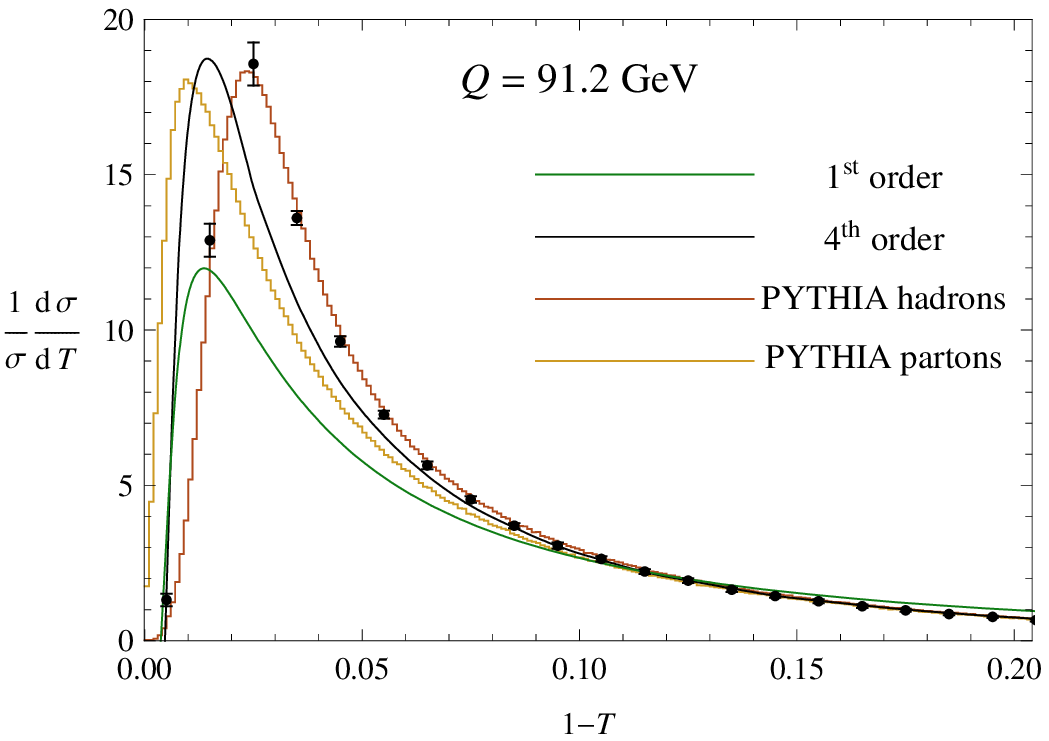} &
\includegraphics[width=0.48\textwidth]{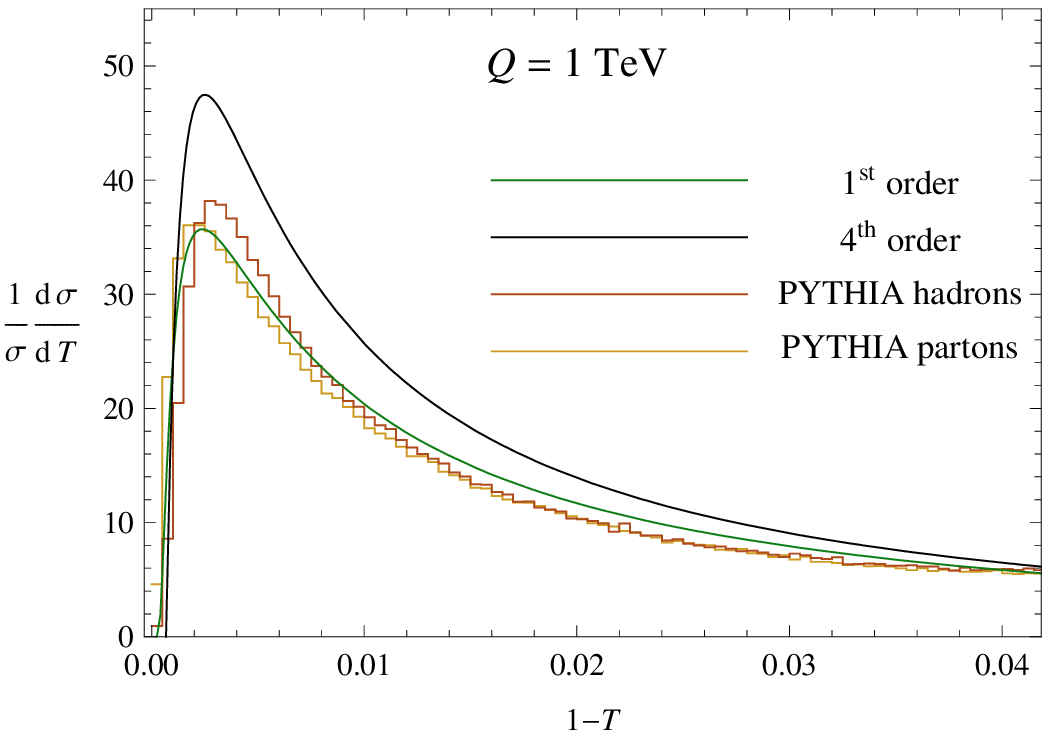}
\end{tabular}
\end{center}
\vspace{-0.6cm}
\caption{Comparison between theoretical predictions in effective field theory at
first order and fourth order, as defined in table \ref{tab:ords}, and {\sc{pythia}} at the parton and hadron level. {\sc{aleph}}  data is included in the first panel.}
\label{fig:pythia}
\end{figure}

As shown in Figure \ref{fig:pythia} {\sc{pythia}} agrees with the
{\sc{aleph}} data better than our 4th order resummed and matched theoretical
calculation. How is this possible in a leading-log shower with leading-order
matrix elements? The answer is that part of what is being tuned to data in the
Monte Carlo program is not just the hadronization model but also some kind of
unfaithful imitation of subleading-log resummation. This is demonstrated
in Figure \ref{fig:pythia}, where {\sc{pythia}} is run at the parton and hadron level
and compared to the 1st order and 4th order resummed matched distributions in
the effective field theory. Even at the parton level, {\sc{pythia}} agrees
more with the 4th order than the 1st order. Moreover, the hadronization
corrections provide something like a shift in the distribution, but cannot
explain the structure of the peak region, which really should be determined by
subleading order resummation.
To demonstrate the danger of trusting a tuned
Monte Carlo generator, we run the same event generator at $Q
= 1$ TeV,  and compare again to the theoretical calculations, see Figure \ref{fig:pythia}. Now
{\sc{pythia}} looks like the leading-order event generator that
it is, and the hadronization corrections are small, but {\sc pythia} undershoots the
more accurate 4th order theoretical prediction. 
At high energy the difference will be more difficult to absorb into non-perturbative
effects since hadronization corrections are small.
One consequence is that these
Monte Carlo generators may be underestimating backgrounds at an ILC by 30\%, and perhaps by a similar magnitude at the LHC as well.

An alternative to correcting the theoretical distribution with a Monte-Carlo
transfer matrix is to include explicitly a theoretical model of non-perturbative corrections 
and then use data to determine its parameters. 
The non-perturbative effects are suppressed by the center-of-mass energy and will scale as 
a power of $\Lambda_{\mathrm{NP}}/Q$, with $\Lambda_{\mathrm{NP}}\sim 1$  GeV 
a scale characteristic of strong-interaction effects. The effective theory analysis shows that since
scales lower than $Q$ appear in the perturbative expansion, there will in fact be power corrections suppressed
by the lowest scale, in this case the soft scale $\tau Q$ which will go as a power of $\Lambda_{\rm NP}/(\tau Q)$. 
For completely inclusive processes first order power corrections are absent, but
one should not expect the leading power to be absent for thrust.

\begin{figure}
\psfrag{x}[]{\small $\alpha_s(m_Z)$}
\psfrag{y}[]{\small $\Lambda_{\mathrm{NP}}$ (GeV)}
\begin{center}
\includegraphics[width=0.8\textwidth]{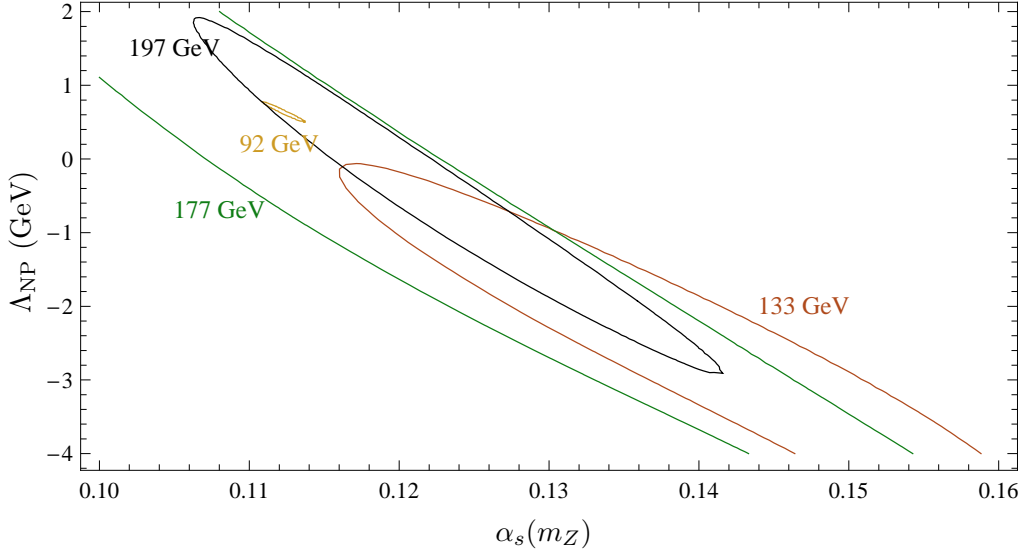}
\end{center}
\vspace{-0.4cm}
\caption{Contours at 95\% confidence level for a fit to the {\sc opal} data of $\alpha_s$ and a non-perturbative
shift parameter $\Lambda_{\rm{NP}}$.}
\label{fig:NPconts}
\end{figure}

The non-perturbative effects will be most important in the soft region for small $\tau$.
The corrections can be parameterized by a non-perturbative shape function which is convoluted with the perturbative soft function \cite{Korchemsky:1998ev,Korchemsky:1999kt}
\begin{equation}
  S (k, \mu) \rightarrow \int \rd k' S (k - k', \mu) S_{\mathrm{NP}} (k', \mu)\,.
\end{equation}
Then one can parametrize $S_{\mathrm{NP}}(k)$ with a few-parameter family of
distributions \cite{Korchemsky:2000kp}. For example, a common model is $S_{\mathrm{NP}} (k) = \delta
(k - \Lambda_{\mathrm{NP}})$, which leads to an overall shift in the thrust
distribution. 
Figure \ref{fig:NPconts} shows the result of a simultaneous fit to
$\Lambda_{\mathrm{NP}}$ and $\alpha_s$ for the {\sc opal} data. 
From this rough analysis one can see that the fit to \lep data has trouble distinguishing the effect of raising the shift parameter 
from increasing the coupling -- both variations increase the theoretical prediction in all bins where the fit to data is performed.

Much of the evidence for a shift in event shape
distributions has come from comparisons to data of calculations done at
NLO or with resummation at NLL \cite{MovillaFernandez:2001ed}. It
would be very interesting to reconsider these analyses including information
from NNLO and with N$^3$LL resummation. 
To extract the soft shape function a detailed analysis, including lower energy data, should be performed. At lower energies the effect of the power corrections will be more pronounced so that the parameters of the shape function can be determined and then used in the extraction of $\alpha_s$ from higher energy data. 
The high statistics {\sc{jade}} data with energies from $Q=22-44$ GeV might be particularly suitable for such an analysis \cite{MovillaFernandez:1997fr}.

In our Monte-Carlo studies, we find that quark-mass effects at \lepone are of order 1\%. They tend to increase $\alpha_s$, while hadronization effects lower the central value. In fixed-order perturbation theory, the quark-mass effects have been evaluated  at 
NLO~\cite{Bernreuther:1997jn,Brandenburg:1997pu,Nason:1997nw,Rodrigo:1997gy}. 
Using the factorization theorem for the production of massive quark jets \cite{Fleming:2007qr} and the recent two-loop result for the massive jet-function \cite{Jain:2008gb}, it is would be possible to perform the resummation also 
for the $b$-quark contribution. Since the quark mass corrections are 
small in the region where we extract $\alpha_s$, a fixed-order treatment might be sufficient.
Additional issues involved in matching the perturbative soft and non-perturbative shape functions were discussed recently in~\cite{Hoang:2007vb}.

Since neither Monte-Carlo hadronization corrections nor a simple non-perturbative shift model are satisfactory, we conclude that the best option at this point is to fit the parton-level distribution. To estimate the hadronization uncertainties, we simply lift the errors from
previous studies of the {\sc{aleph}} and {\sc{opal}} data. Numerically this is essentially equivalent to using {\sc ariadne} to calculate the hadronization and quark-mass corrections and the difference to {\sc pythia} as an estimate of the resulting uncertainty, as can be seen in Tables~\ref{tab:alephresults}~and~\ref{tab:opalresults}. With the increased perturbative precision of our result, it would be important to get better control over hadronization effects and to have a more reliable way to assess the associated uncertainty. As we discussed above, this can be achieved with a dedicated shape-function analysis involving also lower energy data. 

\section{Conclusions}

We have resummed the leading logarithmic corrections to the thrust distribution to N$^3$LL. Our calculation is based on an all-order factorization theorem for the thrust distribution in the two-jet region $T\to 1$. The traditional method for resummation of event shapes is limited to NLL. The present paper goes beyond this not only by one but by two orders in logarithmic accuracy.

The factorization theorem, obtained using Soft-Collinear Effective Theory, separates the contributions associated with different energy scales in a transparent way. 
Those associated with higher energy scales are absorbed into Wilson coefficients.
Solving the renormalization-group equations resums large perturbative logarithms of scale ratios. 
 An advantage of the effective theory treatment is that the factorization theorem is derived at the operator level. The different building blocks in the factorization theorem are given by operator matrix elements and appear in a variety of other processes. With the exception of the two-loop constant in the soft function, all the necessary ingredients to the factorization theorem were known to N$^3$LL accuracy from resummations of other processes. We have determined the missing two-loop constant numerically using 
effective field theory and an NLO fixed-order event generator.

Comparing to fixed-order results, we found that the logarithmically enhanced pieces, determined
by a few constants in the effective theory,
amount to the bulk of the fixed-order results, even away from the endpoint $T\to 1$. 
Of particular interest is the comparison at NNLO. The necessary fixed-order calculation has been completed only recently and so far not been independently checked.
 The close agreement with the logarithmic contributions we derive provides a non-trivial check on both calculations. Once matched to the full fixed-order result, our result is valid not only to N$^3$LL accuracy, but also to NNLO in fixed-order perturbation theory. 
Matching improves our result away from the endpoint region, but numerically the matching corrections are small, in particular at NNLO.

Our result is the most precise calculation of an event shape to date, and we have used it to perform a precision determination of $\alpha_s$ using {\sc{aleph}} and {\sc{opal}} data. Our final combined result is
 \begin{align}
  \alpha_s (m_Z) &= 0.1172 \pm 0.0010 (\mathrm{stat}) \pm 0.0008 (\mathrm{sys}) \pm
  0.0012 (\mathrm{had}) \pm 0.0012 (\mathrm{pert}) \nonumber \\
& = 0.1172 \pm 0.0022 \nonumber\,.
\end{align}
Unfortunately, we had to combine different data sets with the conservative assumption that systematic errors are
completely correlated. We also had to use the fit regions selected by the experiments
which are not optimized to our higher order calculation. 
An improved error analysis would involve information from the collaborations about correlations
which is not
publicly available. 

With the resummed calculation, 
the perturbative uncertainty is finally smaller than the other uncertainties at each energy, 
in contrast to earlier results where the perturbative error was dominant. 
With the reduction in perturbative uncertainty, the hadronization error has become a relatively large contribution
to the total error. To reduce it, one could parameterize the non-perturbative effects with a shape function,
and then extract this shape function from data at \lep and lower energy experiments. In addition,
one could account explicitly for quark mass effects which should help reduce the systematic errors.

Even though the perturbative error is greatly reduced by including resummation, the technique used
to estimate this error may be unduly conservative. 
We have followed the standard procedure and used a collection of scale variations to estimate terms higher
order in $\alpha_s$. An alternative method, which we have suggested here, is to attempt a more sophisticated
guess at the effects that a higher order calculation might have. At one higher order the
resummed distribution is known up to a handful of numbers,
such as higher-loop anomalous dimensions. So we can extrapolate an approximation
to these numbers and use that directly. This procedure results in smaller and perhaps more realistic 
perturbative errors, although we have not used the errors derived this way for the final results.

The effective theory can also be used to study other event shapes. For example, heavy- and light-jet masses 
can be obtained with minimal modifications from the formulae given here~\cite{Schwartz:2007ib,Fleming:2007qr}. 
These observables involve the same hard and jet functions as (\ref{scetfact}) and the necessary soft function 
can be determined in the same way as we did here for thrust. The factorization theorem for a wider class of event shapes, including jet-broadening and the $C$-parameter was derived recently \cite{Bauer:2008dt}. Its form is the same as (\ref{scetfact}), except that it involves different jet-functions which depend on the variable under consideration. 
To reach the same accuracy we have achieved here, additional perturbative calculations will thus be necessary.

We could also try to use the same techniques to calculate precision observables in a hadronic environment. Many of the necessary ingredients have already been understood from the threshold resummation for inclusive processes such as deep-inelastic scattering and Drell-Yan production. Despite the complication of hadronic initial states, a precision calculation of jet-observables relevant for the LHC seems feasible. Considering the discrepancy
we found between {\sc pythia} and the fourth-order effective theory prediction for the thrust distribution at 1 TeV
 (see Figure~\ref{fig:pythia}), having a systematically improvable way to perform resummations might be vital for the LHC.
In addition, given the size of the logarithmic corrections found here, it is likely that
many fixed-order calculations can be improved using methods of effective field theory.

\subsection*{Acknowledgements}
The authors would like to thank Keith Ellis, Thomas Gehrmann, Carlo Oleari, Hasko Stenzel, Morris Swartz, Jan Winter and Giulia Zanderighi. M.D.S.\ is supported in part by the National Science Foundation under grant NSF-PHY-0401513 and by the Johns Hopkins Theoretical Interdisciplinary Physics and Astronomy Center. The research of T.B.\ was supported by the Department of Energy under Grant DE-AC02-76CH03000. Fermilab is operated by the Fermi Research Alliance under contract with the U.S.\ Department of Energy.
\begin{appendix}

\section{Anomalous dimensions\label{sec:anomalous}}

The QCD beta-function satisfies
\begin{align}\label{eqbeta}
  \frac{\mathrm{d}\alpha_s(\mu)}{\mathrm{d}\ln \mu} &= \beta (\alpha_s(\mu)) \,,\\
  \beta (\alpha) &=- 2 \alpha\, \left[ \left( \frac{\alpha}{4 \pi} \right) \beta_0 + \left(
  \frac{\alpha}{4 \pi} \right)^2 \beta_1 + \left( \frac{\alpha}{4 \pi}
  \right)^3 \beta_2 + \cdots \right]\,,
\end{align}
with
\begin{align}
\beta_0 &=\frac{11}{3} C_A - \frac{4}{3} T_F n_f \,, \nonumber \\
\beta_1 &= \frac{34}{3} C_A^2 - \frac{20}{3} C_A T_F n_f - 4 C_F T_F n_f \,,\\
\beta_2 &= \frac{325}{54} n_f^2 - \frac{5033}{18} n_f + \frac{2857}{2}\,, \nonumber \\
\beta_3 &= \frac{1093}{729} n_f^3 
+\left(\frac{50065}{162} + \frac{6472}{81}\zeta_3\right) n_f^2
+\left(-\frac{1078361}{162} - \frac{6508}{27} \zeta_3 \right) n_f + 3564 \zeta_3 + \frac{149753}{6} \nonumber\,,
\end{align}
where we have written $\beta_0$ and $\beta_1$ in terms of the Casimir invariants $C_F=\frac{4}{3}$, $C_A=3$ and $T_F=\frac{1}{2}$, but have evaluated $\beta_2$ and $\beta_3$ for $N=3$ colors. The RG equation (\ref{eqbeta}) has a solution in terms of $L = \ln \frac{\mu^2}{\Lambda^2}$
\begin{multline}
  \alpha_s (\mu) = \frac{4 \pi}{\beta_0} \left[ \frac{1}{L} -
  \frac{\beta_1}{\beta_0^2 L^2} \ln L + \frac{\beta_1^2}{\beta_0^4 L^3}
  (\ln^2 L - \ln L - 1) + \frac{\beta_2}{\beta_0^3 L^3} \right. \\
  \left. + \frac{\beta_1^3}{\beta_0^6 L^4} \left( - \ln^3 L + \frac{5}{2}
  \ln^2 L + 2 \ln L - \frac{1}{2} \right) - 3 \frac{\beta_1
  \beta_2}{\beta_0^5 L^4} \ln L + \frac{\beta_3}{2 \beta_0^4 L^4} \frac{}{}
  \right]\,.
\end{multline}
It is also useful sometimes to work with perturbative expansion of $\alpha_s
(\mu)$ in terms of $\alpha_s$ at a fixed renormalization scale, $\mu_R$:
\begin{multline}
  \alpha_s (\mu) = \alpha_s(\mu_R) - \frac{\alpha_s^2(\mu_R)}{2 \pi} \beta_0 \ln
  \frac{\mu}{\mu_R} + \frac{\alpha_s^3(\mu_R)}{8 \pi^2} \left( - \beta_1 \ln
  \frac{\mu}{\mu_R} + 2 \beta_0^2 \ln^2 \frac{\mu}{\mu_R} \right) \\
  + \frac{\alpha_s^4(\mu_R)}{32 \pi^2} \left( - \beta_2 \ln \frac{\mu}{\mu_R} + 5
  \beta_0 \beta_1 \ln^2 \frac{\mu}{\mu_R} - 4 \beta_0^3 \ln^3
  \frac{\mu}{\mu_R} \right) + \cdots \,.
\end{multline}

We write the perturbative expansion of the anomalous dimensions  as\begin{eqnarray}
  \Gamma_{\mathrm{cusp}} (\alpha) &=& \left( \frac{\alpha}{4 \pi} \right) \Gamma_0
  + \left( \frac{\alpha}{4 \pi} \right)^2 \Gamma_1 + \left( \frac{\alpha}{4
  \pi} \right)^3 \Gamma_2 + \cdots \,,\nonumber\\
  \gamma_H (\alpha) &=& \left( \frac{\alpha}{4 \pi} \right) \gamma_0^H + \left(
  \frac{\alpha}{4 \pi} \right)^2 \gamma_1^H + \left( \frac{\alpha}{4 \pi}
  \right)^3 \gamma_2^H + \cdots\,, \nonumber\\
  \gamma_J (\alpha) &=& \left( \frac{\alpha}{4 \pi} \right) \gamma_0^J + \left(
  \frac{\alpha}{4 \pi} \right)^2 \gamma_1^J + \left( \frac{\alpha}{4 \pi}
  \right)^3 \gamma_2^J + \cdots \,,\nonumber\\
  \gamma^S(\alpha) &=& \gamma^H(\alpha)-2\gamma^J(\alpha) \,.
\end{eqnarray}
The exact anomalous dimensions are known to $\alpha_s^3$.
The anomalous dimensions for the hard function are
\begin{align}
  \gamma_0^H &= - 6 C_F\,, \nonumber \\
  \gamma_1^H &= C_F^2 (- 3 + 4 \pi^2 - 48 \zeta_3) + C_F C_A \left( -
  \frac{961}{27} - \frac{11 \pi^2}{3} + 52 \zeta_3 \right) + C_F T_F n_f
  \left( \frac{260}{27} + \frac{4 \pi^2}{3} \right) \,, \nonumber 
\\   \gamma_2^H &= - 1.856 n_f^2 + 259.3 n_f - 1499 \,.
\end{align}
For the jet function
\begin{align}
  \gamma_0^J &= - 3 C_F \,, \nonumber \\
  \gamma_1^J &= C_F^2 (- \frac{3}{2} + 2 \pi^2 - 24 \zeta_3) + C_F C_A \left( -
  \frac{1769}{54} - \frac{11 \pi^2}{9} + 40 \zeta_3 \right) + C_F T_F
  n_f \left( \frac{242}{27} + \frac{4 \pi^2}{9} \right)\,,  \nonumber \\
  \gamma_2^J &= - 0.7255 n_f^2 + 85.35 n_f - 203.8\, .
\end{align}
For the soft function
\begin{align}
\gamma_0^S &= \gamma_0^H - 2 \gamma_0^J \,,&
\gamma_1^S &= \gamma_1^H - 2 \gamma_1^J \,,& 
\gamma_2^S &= \gamma_2^H - 2 \gamma_2^J\, . \hspace*{5cm}\phantom{a}
\end{align}
And for the cusp anomalous dimension
\begin{align}
  \Gamma_0 & = 4 C_F\,,\hspace*{13cm}\phantom{a} \nonumber\\
  \Gamma_1 &= 4 C_F \left[ C_A \left( \frac{67}{9} - \frac{\pi^2}{3} \right) -
  \frac{20}{9} C_F T_F n_f \right]  \,, \nonumber \\
  \Gamma_2 &= - 0.7901 n_f^2 - 183.2 n_f + 1175 \,,&  \nonumber\\
  \Gamma_3 &\approx \frac{\Gamma_2^2}{\Gamma_1}\,.
\end{align}
Analytical expressions for the three-loop terms $\gamma_2^H$, $\gamma_2^J$ and $\Gamma_2$ can be found in~\cite{Becher:2006mr}. 
The $\alpha_s^4$ part of the cusp anomalous dimension is not known and we estimate it using a Pad\'e approximation. The same approximation works well at $\alpha_s^3$ and in any case our results are very insensitive to the value of $\Gamma_3$.

\section{Hard, jet and soft function\label{sec:Hjs}}
The hard function can be written as 
\begin{equation}
H(Q^2,\mu) = h(\ln\frac{Q^2}{\mu^2},\mu)\,, \label{hardfdef}
\end{equation}
where to three-loop order 
\begin{eqnarray}\label{hardfun}
h(L,\mu)&=&1+\left(\frac{\alpha_s}{4 \pi }\right) \bigg[-\frac{1}{2} \Gamma_0 L^2-\gamma^H_0 L+c^H_1\bigg]
+\left(\frac{\alpha_s}{4 \pi }\right)^2 \bigg[\frac{1}{8} \Gamma
  _0^2 L^4+\bigg(\frac{\beta_0 \Gamma_0}{6}+\frac{\gamma^H_0 \Gamma_0}{2}\bigg)
   L^3 \nonumber \\
 &&\hspace{1cm}  +\bigg(\frac{(\gamma^H_0)^2}{2}+\frac{\beta _0 \gamma^H_0}{2}-\frac{\Gamma_1}{2}\bigg) L^2-\gamma^H_1 L+c^H_1 \bigg(-\frac{L^2 \Gamma _0}{2}+L \left(-\beta _0-\gamma^H
   _0\right)\bigg)+c^H_2\bigg] \nonumber\\
&&   +\left(\frac{\alpha _s}{4 \pi }\right)^3 
   \bigg[-\frac{1}{48} \Gamma _0^3 L^6+\bigg(-\frac{1}{12} \beta _0
   \Gamma _0^2-\frac{1}{8} \gamma^H _0 \Gamma _0^2\bigg) L^5\nonumber\\
   &&\hspace{2cm}   +\bigg(-\frac{1}{12} \Gamma
   _0 \beta _0^2-\frac{5}{12} \gamma^H _0 \Gamma _0 \beta _0    -\frac{1}{4} (\gamma_0^H)^2
   \Gamma _0+\frac{\Gamma _0 \Gamma _1}{4}\bigg) L^4 \nonumber\\
&&\hspace{2cm}  +\bigg(-\frac{(\gamma^H_0)^3}{6}-\frac{1}{2} \beta _0 (\gamma^H_0)^2-\frac{1}{3} \beta _0^2 \gamma^H_0+\frac{\Gamma _1 \gamma^H_0}{2}+\frac{\beta _1 \Gamma _0}{6}+\frac{\gamma^H_1
   \Gamma _0}{2}+\frac{\beta _0 \Gamma _1}{3}\bigg) L^3 \nonumber\\
 &&\hspace{2cm}   +\bigg(\frac{\beta _1 \gamma^H_0}{2}+\gamma^H_1 \gamma^H_0+\beta _0 \gamma^H_1-\frac{\Gamma _2}{2}\bigg) L^2-\gamma^H_2 L    \nonumber \\
&& \hspace{2cm}  +c^H_1 \bigg\{\frac{1}{8} \Gamma _0^2 L^4+\bigg(\frac{2 \beta _0 \Gamma
   _0}{3}+\frac{\gamma^H_0 \Gamma _0}{2}\bigg) L^3 
   +\bigg(\beta _0^2+\frac{3 \gamma^H_0
   \beta _0}{2}+\frac{(\gamma^H_0)^2}{2}-\frac{\Gamma _1}{2}\bigg) L^2 \nonumber\\ 
     &&\hspace{3cm} +\left(-\beta
   _1-\gamma^H_1\right) L\bigg\}    +c^H_2 \bigg\{L \left(-2 \beta _0-\gamma^H_0\right)-\frac{L^2 \Gamma
   _0}{2}\bigg\}+c^H_3\bigg]\,.
   \end{eqnarray}
 The three-loop constant $c_3^H$ is not yet known but only contributes to the $\delta(\tau)$ part of the thrust distribution. The values of the lower order constants are
\begin{align}
  c_1^H & = C_F \left(-16+\frac{7 \pi ^2}{3}\right) \,,\nonumber \\
  c_2^H &= C_F^2 \left(\frac{511}{4}-\frac{83 \pi
   ^2}{3}+\frac{67 \pi ^4}{30}-60 \zeta_3  \right) + C_F C_A \left(-\frac{51157}{324}+\frac{1061 \pi ^2}{54}-\frac{8 \pi
   ^4}{45}+\frac{626 \zeta_3}{9}  \right) \nonumber \\
& \hspace{1cm}  + C_F T_F n_f \left( \frac{4085}{81}-\frac{182 \pi ^2}{27}+\frac{8 \zeta_3}{9} \right) \,.
\end{align}
   
The expression for $H(Q,\mu)$ is obtained by solving the RG-equation (\ref{Hrge}) order by order in $\alpha _s$. The RG equations for the Laplace transformed jet function and soft function have the same form so that their explicit forms are obtained from the above result using simple substitution rules. Defining the Laplace transform of the cross section as
\begin{equation}\label{laplace}
\widetilde t(\nu)
   = \int_0^\infty\!\rd\tau\,e^{-\nu \tau}\,
   \frac{1}{\sigma_0} \frac{\rd\sigma}{\rd\tau} \,, 
\end{equation}
the cross section factors into the product of the Laplace transforms of the jet- and soft functions:
\begin{equation}\label{factLaplace}
\widetilde t(\nu)= H(Q^2,\mu)\, \left[\widetilde j\Big(\ln\frac{ s Q^2}{\mu^2},\mu\Big)\right]^2\, \widetilde s_T\Big( \ln\frac{s Q}{\mu},\mu\Big),  \,
\quad s\equiv \frac{1}{e^{\gamma_E} \nu}.
\end{equation}
The Laplace transforms  $\widetilde j$ and $\widetilde s_T$ of the jet and soft functions are defined as in (\ref{laplace}). After writing these as functions of a logarithm of the argument, the RG equations simplify to
\begin{eqnarray}\label{gammaV}
\label{jtildeevol}
   \frac{\rd}{\rd\ln\mu}\,\widetilde j\Big(\ln\frac{s Q^2}{\mu^2},\mu \Big)
   &=& \left[- 2\Gamma_{\rm cusp}(\alpha_s)\,\ln\frac{s Q^2}{\mu^2}
   - 2\gamma^J(\alpha_s) \right]
   {\widetilde j}\Big(\ln\frac{s Q^2}{\mu^2},\mu \Big) \,, \\
   \label{stildeevol}
   \frac{\rd}{\rd\ln\mu}\,\widetilde s_T\Big(\ln\frac{s Q}{\mu},\mu \Big)
  & = & \left[ 4\Gamma_{\rm cusp}(\alpha_s)\,\ln\frac{s Q}{\mu}
   - 2\gamma^S(\alpha_s) \right]
   {\widetilde s}_T\Big(\ln\frac{s Q}{\mu},\mu \Big) \,.
\end{eqnarray} 
Comparing to the RG equation for the hard function Eq. (\ref{Hrge}), and looking at Eq.~\eqref{hardfdef}
one sees that the expression for the jet-function $\widetilde j(L,\mu)$ is obtained from (\ref{hardfun}), by simple substitutions:
\begin{eqnarray}
\widetilde{j}(L,\mu) &=& h(L,\mu) \quad \mathrm{with} \quad 
\gamma^H \to -\gamma^J,\,c^H \to c^J, \,\,\mathrm{and}\,\, \Gamma_{\rm cusp} \to -\Gamma_{\rm cusp},\,\\
\widetilde{s}_T(L,\mu) &=& h(2 L,\mu) \quad \mathrm{with} \quad \gamma^H \to -\gamma^S,\, \,\mathrm{and}\,\,
c^H \to c^S \,.
\end{eqnarray}
The constants for the jet and soft functions are
\begin{align}
  c_1^J &= C_F \left( 7 - \frac{2 \pi^2}{3} \right) \nonumber \,,\\
  c_2^J &= C_F^2 \left( \frac{205}{8} - \frac{97 \pi^2}{12} + \frac{61
  \pi^4}{90} - 6 \zeta_3 \right) + C_F C_A \left( \frac{53129}{648} -
  \frac{155 \pi^2}{36} - \frac{37 \pi^4}{180} - 18 \zeta_3 \right) \nonumber\\
 &\hspace{1cm} + C_F T_F n_f  \left( - \frac{4057}{162} + \frac{13 \pi^2}{9} \right)\,,\\
  c_1^S &= - C_F \pi^2  \,, \nonumber\\
  c_2^S &=  C_F^2 c_{2_{CF}}^S + C_F C_A c_{2,{C_A}}^S + 
  C_FT_F\,n_f c_{2,{n_f}}^S  \,.\nonumber
\end{align}
The constants $c_{2_{CF}}^S, c_{2_{CA}}^S$ and
$c_{2_{n_f}}^S$ are extracted numerically as explained in Section \ref{sec:match}. We found
\begin{align}
  c_{2_{CF}}^S &= 58\pm2, & c_{2_{CA}}^S &= -
  60\pm 1, & c_{2_{n_f}}^S = 43 \pm 1 \,.
\end{align}

\section{Singular terms in the thrust distribution\label{sec:singular}}

The fixed-order distributions can be written in terms of delta functions and
plus distributions.
\begin{equation}
  D (\tau) = \delta (\tau)\, D_{\delta} (\tau) 
+ \left( \frac{\alpha_s}{2 \pi}  \right) \left[ D_A (\tau) \right]_{+} 
+ \left( \frac{\alpha_s}{2 \pi}  \right)^2 \left[ D_B (\tau) \right]_{+} 
+ \left( \frac{\alpha_s}{2 \pi}  \right)^3 \left[ D_C (\tau) \right]_{+} \, .
\end{equation}
The delta-function terms are known to $\alpha_s^2$ accuracy
 \begin{align}\label{deltaterms}
  D_{\delta}  &= 1 + \left( \frac{\alpha_s}{4 \pi} \right)C_F
  \left( - 2 + \frac{2 \pi^2}{3} \right)  
  + \left( \frac{\alpha_s}{4 \pi} \right)^2 \left\{ C_F^2 \left( 4 - \frac{3
  \pi^2}{2} + \frac{\pi^4}{18} - 24 \zeta_3 + c_{2_{CF}}^S \right) 
  \right.\quad \\
 &  \left. + C_A C_F \left( \frac{493}{81} + \frac{85 \pi^2}{6} - \frac{73 \pi^4}{90} +
  \frac{566 \zeta_3}{9} + c_{2_{CA}}^S \right)  + C_F T_F n_f \left( \frac{28}{81} - \frac{14 \pi^2}{3} -
  \frac{88 \zeta_3}{9} + c_{2_{nf}}^S \right) \right\} \,. \nonumber
\end{align}
Our results allow us to derive all plus-distribution terms to $\alpha_s^3$. We find
{ \begin{align}
  D_A (\tau) &= \frac{1}{\tau}\left\{C_F \left[- 4 \ln \tau -3\right]\right\}\, , \nonumber \\
  D_B (\tau) &= \frac{1}{\tau} \left\{ C_F^2 \left[ 8 \ln^3 \tau + 18 \ln^2
  \tau + (13 - 4 \pi^2) \ln \tau + \frac{9}{4} - 2 \pi^2 + 4 \zeta_3 \right] \right. \\
& \hspace{1cm}  \left. + C_F T_F n_f \left[ - 4 \ln^2 \tau + \frac{22}{9} \ln \tau
  + 5 \right] \right. \nonumber \\ 
  &\left. \hspace{1cm}+ C_F C_A \left[ 11 \ln^2 \tau + (- \frac{169}{18} + \frac{2 \pi^2}{3}) \ln \tau 
- \frac{57}{4} + 6 \zeta_3 \right] \right\} \, ,\nonumber \\
D_C (\tau) &= \frac{1}{\tau}\bigg\{ C_F^3 \bigg[-8 \ln^5\tau-30 \ln^4\tau +\ln^3\tau \left(-44+\frac{40 \pi ^2}{3}\right)+\ln^2\tau \big(-88 \zeta _3+24 \pi ^2\nonumber
 \\
 & -27\big) + \ln\tau \left(-c_{2_{C_F}}^S-96 \zeta _3-\frac{53 \pi ^4}{90}+\frac{79 \pi ^2}{6}-\frac{17}{2}\right) +16 \pi ^2 \zeta _3-39 \zeta _3-132 \zeta _5\nonumber
 \\
 &+\frac{19 \pi ^4}{120} +\frac{5}{8} \pi ^2-\frac{47}{8}-\frac{3}{4} c_{2_{C_F}}^S \bigg] +C_F^2 n_f T_F \bigg[\frac{40 \ln^4\tau}{3}+\frac{56 \ln^3\tau}{9}+\ln^2\tau \left(-43-\frac{28 \pi ^2}{3}\right) \nonumber
  \\
& +\ln\tau \left(- c_{2_{n_f}}^S+\frac{664 \zeta _3}{9}+\frac{164 \pi ^2}{27}-\frac{1495}{81}\right)+\frac{274 \zeta _3}{9}-\frac{31 \pi ^4}{45}+\frac{56 \pi ^2}{9}+\frac{1511}{108} \nonumber
\\
& + \frac{2}{3}c_{2_{C_F}}^S-\frac{3}{4} c_{2_{n_f}}^S \bigg] + C_F n_f^2 T_F^2 \bigg[-\frac{112 \ln^3\tau}{27}+\frac{68 \ln^2\tau}{9}+\ln\tau \left(\frac{140}{81}+\frac{16 \pi ^2}{27}\right) \nonumber\\
&-\frac{176 \zeta _3}{27}-\frac{64 \pi ^2}{81}-\frac{3598}{243}+\frac{2}{3} c_{2_{n_f}}^S \bigg] + C_F C_A^2 \bigg[-\frac{847 \ln^3\tau}{27}+\ln^2\tau \left(\frac{3197}{36}-\frac{11 \pi ^2}{3}\right)  \nonumber 
\\
&+\ln\tau \left(22 \zeta _3-\frac{11 \pi ^4}{45}+\frac{85 \pi ^2}{9}-\frac{11323}{324}\right)-10 \zeta _5+\frac{361 \zeta _3}{27}+\frac{541 \pi ^4}{540}-\frac{739 \pi ^2}{81}\nonumber
\\
& -\frac{77099}{486}-\frac{11}{6} c_{2_{C_A}}^S \bigg]  
+ C_F^2 C_A \bigg[-\frac{110 \ln^4\tau}{3}+\ln^3\tau \left(-\frac{58}{9}-\frac{8 \pi ^2}{3}\right)
+\ln^2\tau \left(-36 \zeta _3 \nonumber 
\right. \\ 
& \left.+\frac{68 \pi ^2}{3}+\frac{467}{4}\right)+\ln\tau \left(-\frac{2870 \zeta _3}{9}+\frac{173 \pi ^4}{90}-\frac{625 \pi ^2}{27}+\frac{29663}{324}- c_{2_{C_A}}^S\right)-30 \zeta _5 \nonumber 
\\
& -\frac{1861 \zeta _3}{18}+\frac{973 \pi ^4}{360}-\frac{317 \pi ^2}{18}-\frac{49}{27}-\frac{11}{6} c_{2_{C_F}}^S-\frac{3}{4} c_{2_{C_A}}^S\bigg] +C_A C_F n_f T_F \bigg[\frac{616 \ln^3 \tau}{27} \nonumber \\
&+\ln^2\tau \left(\frac{4 \pi ^2}{3}-\frac{512}{9}\right)+\ln\tau \left(8 \zeta _3-\frac{128 \pi ^2}{27}+\frac{673}{81}\right)+\frac{608 \zeta_3}{27}-\frac{10 \pi ^4}{27}+\frac{430 \pi ^2}{81} \nonumber 
\\
& +\frac{24844}{243} -\frac{11}{6} c_{2_{n_f}}^S+\frac{2}{3} c_{2_{C_A}}^S \bigg] \bigg\}\,.  \nonumber
\end{align}}
The numerical values of $c_{2_{C_F}}^S$, $c_{2_{C_A}}^S$ and $c_{2_{n_f}}^S$ were given in  (\ref{softConst}).
\begin{table}[t!]
\begin{center}
\begin{tabular}{|c|c|c|c|}\hline
& $\alpha_s$ & $\alpha_s^2$ & $\alpha_s^3$ \\ \hline
\multirow{2}{*}{LL} & $G_{12}$ & $G_{23}$ & $G_{34}$  \\
& -2.667 & -10.22 & -45.72\\ \hline
\multirow{2}{*}{NLL} & $G_{11}$ &  $G_{22}$ & $G_{33}$  \\
 & 4  & -24.94 & -285.1 \\ \hline
\multirow{2}{*}{N$^2$LL} & $C_1$ & $G_{21}$ & $G_{32}$  \\ 
& 1.053 &  21.82  & -230.7 \\ \hline
\multirow{2}{*}{N$^3$LL} & & $C_2$ & $G_{31}$   \\
& -- & $73.\pm 2. $ &  $293. \pm 24.$    \\ \hline
\end{tabular}
\end{center}
\caption{Numerical values for the expansion coefficients of  $R(\tau)$ as defined in (\ref{Rexpand}). \label{GijNumerical}}
\end{table}

To compare with the existing literature and for the readers convenience, we also quote the third order result for the quantity $R(\tau)$, which is is often written in the form
\begin{equation}\label{Rexpand}
R(\tau) =\int_0^\tau \frac{1}{\sigma_{\rm had}} \frac{d\sigma}{d\tau} =  \left(1+\sum_{k=1}^{\infty} C_k \left(\frac{\alpha_s}{2\pi}\right)^k\right) \exp\left[ \sum_{i=1}^\infty \sum_{j=0}^{i+1} \left(\frac{\alpha_s}{2\pi}\right)^i \ln^j\frac{1}{\tau}  G_{ij}\right] \,.
\end{equation}
We normalize here to the total hadronic cross section $\sigma_{\rm had}$ given in (\ref{sigtot}) instead of the Born cross section $\sigma_0$. Our result provides the normalization of $R(\tau)$ to second order
\begin{align}\label{Cicoeff}
C_1 =& C_F \left(-\frac{5}{2}+\frac{\pi ^2}{3}\right) \, ,\nonumber \\
C_2 =& C_F^2 \left(-6 \zeta (3)+\frac{\pi ^4}{72}-\frac{7 \pi^2}{8}+\frac{41}{8}+\frac{c_{2_{C_F}}^S}{4} \right) +C_A C_F \left( \frac{481 \zeta (3)}{18}-\frac{73 \pi^4}{360}+\frac{85 \pi ^2}{24}-\frac{8977}{648} \right.  \nonumber \\ 
& \left.+\frac{c_{2_{C_A}}^S}{4}  \right)
 + C_F n_f T_F \left(-\frac{58 \zeta (3)}{9}-\frac{7 \pi^2}{6}+\frac{905}{162}+\frac{c_{2_{n_f}}^S}{4} \right)\,,
\end{align}
 and determines all logarithmic terms up to $\alpha_s^3$:
\begin{align}\label{Gijcoeff}
G_{12} &= -2 C_F \, ,\;\;\;  G_{11} = 3 C_F \, ,\nonumber\\ 
G_{23} &= C_F \left(n_f T_F \frac{4 }{3}-C_A \frac{11 }{3}\right) ,\;\;\;
G_{22} = C_F \left(-C_F\frac{4 \pi ^2 }{3}+n_f T_F\frac{11 }{9}+C_A \left(-\frac{169}{36}+\frac{\pi ^2}{3}\right) \right) \, , \nonumber\\ 
G_{21}& = C_F \left(C_F \left(-4 \zeta _3+\pi ^2+\frac{3}{4}\right)-5 n_f T_F+C_A \left(\frac{57}{4}-6 \zeta _3\right)\right)\, , \nonumber\\ 
G_{34}& = C_F \Bigg(-C_A^2\frac{847 }{108}+ C_A n_f T_F\frac{154}{27} -n_f^2 T_F^2 \frac{28}{27} \Bigg) \, , \nonumber \\
G_{33}& = C_F \Bigg(C_A^2 \left(-\frac{3197}{108}+\frac{11 \pi ^2}{9}\right) +n_f T_F C_A \left(\frac{512}{27}-\frac{4 \pi ^2}{9}\right) - n_f^2 T_F^2\frac{68}{27} +   \nonumber\\ 
& C_F n_f T_F \left(2+\frac{8 \pi ^2}{3}\right)- C_F C_A \frac{22 \pi ^2}{3}+C_F^2 \frac{64}{3}  \zeta _3\Bigg) \, ,\nonumber\\
G_{32}& = C_F \Bigg(C_A^2 \left(11 \zeta _3-\frac{11 \pi ^4}{90}+\frac{85 \pi ^2}{18}-\frac{11323}{648}\right) +C_A n_f T_F \left(4 \zeta _3-\frac{64 \pi ^2}{27}+\frac{673}{162}\right) 
 \nonumber \\ & 
 +n_f^2 T_F^2 \left(\frac{70}{81}+\frac{8 \pi ^2}{27}\right) +C_F^2 \left(\frac{8 \pi ^4}{45}-48 \zeta _3\right)+ C_F C_A \left(-110 \zeta _3+\frac{4 \pi ^4}{9}-\frac{70 \pi ^2}{27}+\frac{11}{8}\right)  \nonumber\\ 
&  + C_F n_f T_F \left(32 \zeta _3+\frac{8 \pi ^2}{27}+\frac{43}{6}\right)\Bigg)  \, ,\nonumber\\
G_{31}& = C_F \Bigg( C_F^2 \left(-\frac{44}{3} \pi ^2 \zeta _3+53 \zeta _3+132 \zeta _5-\frac{8 \pi ^4}{15}+\frac{5 \pi ^2}{4}+\frac{29}{8}\right) 
\nonumber\\
& + C_F n_f T_F \left(-\frac{2}{3} c_{2_{C_F}}^S-\frac{208 \zeta _3}{9}+\frac{31 \pi ^4}{45}-\frac{19 \pi ^2}{18}-\frac{77}{4}\right)
 \nonumber\\
&+C_F C_A \left(\frac{11 c_{2_{C_F}}^S}{6}+2 \pi ^2 \zeta _3+\frac{452 \zeta _3}{9}+30 \zeta _5-\frac{377 \pi ^4}{180}+\frac{161 \pi ^2}{72}+\frac{23}{2} \right)
\nonumber\\
&+C_A^2 \left(\frac{11}{6} c_{2_{C_A}}^S-\frac{361 \zeta _3}{27}+10 \zeta _5-\frac{541 \pi ^4}{540}+\frac{739 \pi ^2}{81}+\frac{77099}{486}\right) 
\nonumber\\
& +C_A n_f T_F \left(\frac{11 c_{2_{n_f}}^S}{6}-\frac{2}{3} c_{2_{C_A}}^S-\frac{608 \zeta _3}{27}+\frac{10 \pi ^4}{27}-\frac{430 \pi ^2}{81}-\frac{24844}{243}\right) 
\nonumber\\ 
&+n_f^2 T_F^2 \left(-\frac{2}{3} c_{2_{n_f}}^S+\frac{176 \zeta _3}{27}+\frac{64 \pi ^2}{81}+\frac{3598}{243}\right)  \Bigg) \, .
\end{align}
The numerical values of the above coefficients are listed in Table \ref{GijNumerical}. The NLL coefficients up to $O(\alpha_s^3)$ were given in \cite{Catani:1992ua} and we completely agree with their results. In the same reference the remaining $\alpha_s^2$ coefficients were determined using a fit to the numerical fixed order result with the result $C_1=34\pm 22 $ and $G_{21}=30\pm 10$. Our analytical result agrees with the extracted value of $G_{21}$, but our value of $C_1$ is about a factor of two larger. This disagreement is perhaps not that surprising, given that  \cite{Catani:1992ua} had to extract $C_2$ and $G_{21}$ numerically using a simultaneous fit to both quantities at small $\tau$, where the result is dominated by the contribution from the logarithmic term proportional to $G_{21}$. Since we have the analytical result for $G_{21}$, we are able to extract $C_2$ with much higher precision.

\end{appendix}

\end{document}